\documentclass[longauth]{aa}
\makeatletter
\renewcommand*\aa@pageof{, page \thepage{} of \pageref*{LastPage}}
\makeatother

\usepackage{graphicx}
\usepackage{txfonts}
\usepackage{booktabs}
\usepackage{natbib,twoopt}
\usepackage[breaklinks=true]{hyperref} 
\usepackage{color}
\usepackage{amsmath}

\usepackage{xspace}

\newcommand{\thisgrb}{GRB~150309A\xspace}
\newcommand{\fermi}{{\em Fermi}\xspace}
\newcommand{\kw}{{\em Konus}-Wind\xspace}

\newcommand{\swiftT}{{T$_{\rm sw,0}$}\xspace}
\newcommand{\fermiT}{{T$_{0}$}\xspace}
\newcommand{\keV}{{\rm keV}\xspace}

\newcommand{\swift}{{\em Swift}\xspace}
\newcommand{\tninty}{{T$_{90}$}\xspace}
\newcommand{\mvts}{{t$_{\rm mvts}$}\xspace}

\newcommand{\Ep}{E$_{\rm p}$\xspace}
\newcommand{\sw}[1]{\texttt{#1}}
\graphicspath{{./}{figures/}}
\bibpunct{(}{)}{;}{a}{}{,} 

\begin{document} 

\title{Revealing characteristics of dark \thisgrb: dust extinguished or high-$z$?}
\titlerunning{\thisgrb: A dark burst}
 
\author{A. J. Castro-Tirado\inst{\ref{inst1},\ref{inst2}}
\and Rahul Gupta\inst{\ref{inst3},\ref{inst4}}
\and S. B. Pandey\inst{\ref{inst3}}
\and A. Nicuesa Guelbenzu\inst{\ref{inst5}}
\and S. Eikenberry\inst{\ref{inst6}} 
\and K. Ackley\inst{\ref{inst6}} 
\and A. Gerarts\inst{\ref{inst7}} 
\and A. F. Valeev\inst{\ref{inst8},\ref{inst9}}
\and S. Jeong\inst{\ref{inst1},\ref{inst10}}
\and I. H. Park\inst{\ref{inst10}}
\and S. R. Oates\inst{\ref{inst11}}
\and B.-B. Zhang\inst{\ref{inst12},\ref{inst13}}
\and R. S\'anchez-Ram\'{\i}rez\inst{\ref{inst1},\ref{inst14}}
\and A. Mart\'{\i}n-Carrillo\inst{\ref{inst15}}
\and J. C. Tello\inst{\ref{inst1}} 
\and M. Jel\'{\i}nek\inst{\ref{inst16}}
\and Y.-D. Hu\inst{\ref{inst1},\ref{inst17}}
\and R. Cunniffe\inst{\ref{inst1}}
\and V. V. Sokolov\inst{\ref{inst8}}
\and S. Guziy\inst{\ref{inst1},\ref{inst19},\ref{inst20}}
\and P. Ferrero\inst{\ref{inst1}}
\and M.~D. Caballero-Garc\'{i}a\inst{\ref{inst1}}
\and A. K. Ror\inst{\ref{inst3}}
\and A. Aryan\inst{\ref{inst3},\ref{inst4}}
\and M. A. Castro Tirado\inst{\ref{inst1}}
\and E. Fern\'andez-Garc\'{\i}a\inst{\ref{inst1}}
\and M. Gritsevich\inst{\ref{inst21},\ref{inst22}}
\and I. Olivares\inst{\ref{inst1}}
\and I. P\'erez-Garc\'{\i}a\inst{\ref{inst1}}
\and J. M. Castro Cer\'on\inst{\ref{inst23}}
\and J. Cepa\inst{\ref{inst24}}
}
\institute{Instituto de Astrof\'isica de Andaluc\'ia (IAA-CSIC), Glorieta de la Astronom\'ia s/n, E-18008, Granada, Spain \label{inst1} \\ \email{ajct@iaa.es}
\and
Departamento de Ingenier\'ia de Sistemas y Autom\'atica, Escuela de Ingenier\'ias, Universidad de M\'alaga, C\/. Dr. Ortiz Ramos s\/n, E-29071, M\'alaga, Spain \label{inst2}
\and Aryabhatta Research Institute of Observational Sciences (ARIES), Manora Peak, Nainital-263002, India \label{inst3} \\ \email{rahulbhu.c157@gmail.com}
\and Department of Physics, Deen Dayal Upadhyaya Gorakhpur University, Gorakhpur-273009, India \label{inst4}
\and Th\~{u}ringer Landessternwarte Tautenburg, Sternwarte 5, 07778 Tautenburg, Germany \label{inst5}
\and Department of Astronomy, University of Florida, Bryant Space Science Center, Gainesville, FL 32611, USA\label{inst6} 
\and Grantecan, Santa Cruz de la Palma, Tenerife, Spain. \label{inst7} 
 \and Special Astrophysical Observatory of Russian Academy of Sciences, Nizhniy Arkhyz 369167, Russia \label{inst8}
 \and Crimean Astrophysical Observatory, Russian Academy of Sciences, Nauchnyi 298409, Russia \label{inst9}
\and Institute for Science and Technology in Space, SungKyunKwan University, Suwon 16419, Korea \label{inst10}
 \and School of Physics and Astronomy, University of Birmingham, Birmingham B15 2TT, UK \label{inst11}
 \and School of Astronomy and Space Science, Nanjing University, Nanjing 210093, China \label{inst12} 
 \and Key Laboratory of Modern Astronomy and Astrophysics (Nanjing University), Ministry of Education, Nanjing 210093, China \label{inst13}
 \and INAF, Istituto di Astrofisica e Planetologia Spaziali, via Fosso del Cavaliere 100, I-00133 Rome, Italy \label{inst14}
 \and School of Physics, O’Brien Centre for Science North, University College Dublin, Belfield, Dublin 4, Ireland \label{inst15}
\and Astronomical Institute of the Czech Academy of Sciences (ASU-CAS), Fri\v{c}ova 298, 251 65 Ond\v{r}ejov, CZ \label{inst16}
\and Universidad de Granada, Facultad de Ciencias Campus Fuentenueva s/n, E-18071 Granada, Spain \label{inst17}
\and Nikolaev National University, Nikolska 24, Nikolaev 54030, Ukraine \label{inst19}
\and Nikolaev Astronomical Observatory, Nikolaev 54030, Ukraine \label{inst20}
\and Department of Physics, University of Helsinki, Gustav H\"allstr\"omin katu 2, FI-00014 Helsinki, Finland \label{inst21}
\and Institute of Physics and Technology, Ural Federal University, Mira street 19, 620002 Ekaterinburg, Russia \label{inst22}
\and European Space Astronomy Centre (ESA-ESAC), Camino bajo del Castillo, s/n, Villafranca del Castillo, E-28692 Villanueva de la Ca\~{n}ada (Madrid), Spain \label{inst23}
\and Instituto de Astrof\'{\i}sica de Canarias, Via L\'actea s/n, La Laguna (Tenerife), Spain. \label{inst24}}

\date{Received 2023; accepted}

\abstract
{Dark Gamma-Ray Bursts (GRBs) constitute a significant fraction of the GRB population. In this paper, we present the multiwavelength analysis (both prompt emission and afterglow) of an intense (3.98 $\times$ 10$^{-5}$ erg cm$^{-2}$ using \fermi-Gamma-Ray Burst Monitor) two-episodic \thisgrb observed early on to $\sim$ 114 days post-burst. Despite the strong gamma-ray emission, no optical afterglow was detected for this burst. However, we discovered near-infrared (NIR) afterglow ($K_{\rm S}$-band), $\sim$ 5.2 hours post burst, with the CIRCE instrument mounted at the 10.4m Gran Telescopio Canarias (hereafter GTC).}
{We aim to examine characteristics of \thisgrb as a dark burst and to constrain other properties using multiwavelength observations.}
{We used \fermi observations of \thisgrb to understand the prompt emission mechanisms and jet composition. We performed the early optical observations using the BOOTES robotic telescope and late-time afterglow observations using the GTC. A potential faint host galaxy is also detected at optical wavelength using the GTC. We modelled the potential host galaxy of \thisgrb in order to explore the environment of the burst.}
{The time-resolved spectral analysis of \fermi data indicates a hybrid jet composition consisting of a matter-dominated fireball and magnetic-dominated Poynting flux. GTC observations of the afterglow revealed that the counterpart of \thisgrb was very red, with H-$K_{\rm S}$ $>$ 2.1 mag (95 $\%$ confidence). The red counterpart was not discovered in any bluer filters of \swift UVOT/BOOTES, indicative of high redshift origin. This possibility was discarded based on multiple arguments, such as spectral analysis of X-ray afterglow constrain $z <$ 4.15 and a moderate redshift value obtained using spectral energy distribution (SED) modelling of the potential galaxy. The broadband (X-ray to NIR bands) afterglow SED implies a very dusty host galaxy with deeply embedded GRB (suggesting $A_{\rm V}$ $\gtrsim$ 35 ~mag).} 
{The environment of \thisgrb demands a high extinction towards the line of sight, demanding dust obscuration is the most probable origin of optical darkness and the very red afterglow of \thisgrb. This result makes \thisgrb the highest extinguished GRB known to date.}

\keywords{Gamma-ray burst: general -- Gamma-ray burst: individual: \thisgrb -- techniques: photometric--dust, extinction}

\maketitle

\twocolumn
\section{Introduction}
\label{intro}

Gamma-ray bursts (GRBs) are the most powerful and sudden explosions of gamma-ray radiation, occurring at cosmological distances \citep{2004RvMP...76.1143P, 2015PhR...561....1K}. GRBs can generally be categorised into two families: long-duration (lasting more than two seconds) and short-duration (lasting less than two seconds) burst \citep{kou93}. Long-duration GRBs are believed to originate from the collapse of massive stars (collapsar model). The core collapse of a massive star leads to the formation of a black hole or a neutron star and the succeeding release of a powerful and ultra-relativistic GRB jet \citep{1999ApJ...524..262M, 2006ARA&A..44..507W, 2003ApJ...591..288H, 2020MNRAS.492.4613O}. Additionally, the observational evidence of their progenitors comes from late-time multi-band observations of their afterglows and host galaxies. Some nearby long GRBs have been observed with the photometric and spectroscopic evidence of supernova signatures \citep{2003ApJ...591L..17S, 2017AdAst2017E...5C}, further supporting the collapsar model and a dense environment. Long-duration GRBs are generally found in star-forming regions of their host galaxies \citep{1998ApJ...507L..25B}. On the other hand, Short-duration GRBs are believed to originate from the merger of compact objects and are generally found in older and less active host galaxies \citep{2002ApJ...570..252P, 2002ApJ...571..394B, Abbott_2017, 2017ApJ...848L..14G}. However, in recent times, some of the hybrid events (long-duration GRB from merger and short-duration GRB from collapsar origin) were observed, which further challenges our classical understanding \citep{2006Natur.444.1050D, 2021NatAs...5..917A, 2021NatAs...5..911Z, 2022Natur.612..228T, 2022Natur.612..223R, 2022Natur.612..232Y, 2022ApJ...932....1R}.

There is a subclass of GRBs afterglows, known as `dark bursts' for which no optical counterpart is detected in the face of deep follow-up (fainter than $\sim$ 21-22 mag) in $<$ 24 hours after the trigger, the number of similar sources has been increasing in the present era of GRB research \citep{Jakobsson04, 2011A&A...526A..30G}. These days, dark bursts appear to make up a significant part of the overall GRB population. From 1997 to 2021, about 66 \% (1526/2311) of well-localised bursts are observed with an X-ray afterglow; however, only 38 \% (874/2311) bursts are observed with an optical afterglow\footnote{\url{https://www.mpe.mpg.de/~jcg/grbgen.html}}. In some cases, optical afterglows could not be detected due to a lack of early observations \citep{2003A&A...408L..21P}. But in other cases, deep and early observations reveal that true dark GRBs indeed exist and have a typical fraction of around 25-40 $\%$ of all the bursts \citep{Perleya16}. In the case of \thisgrb, an X-ray counterpart of the burst was discovered by \swift X-ray telescope \citep[XRT;][]{2005SSRv..120..165B} but no optical afterglow was discovered in spite of deep search using the Burst Observer and Optical Transient Exploring System (BOOTES) and Gran Telescopio Canarias (GTC) telescopes. However, observations in the near-infrared (NIR) bands using the Canarias InfraRed Camera Experiment (CIRCE) instrument mounted on the GTC telescope reveal the detection of a very red afterglow. These observations suggest that \thisgrb belongs to the `dark bursts' subclass. 

Several methods exist in the literature to classify these dark GRBs. At the early stage of their studies, GRBs without an optical afterglow were classified as dark GRBs \citep{1998ApJ...493L..27G}. In later studies \citep{DePasquale03, Jakobsson04}, the analogy between the X-ray and optical/near-IR afterglow properties were used for their classification. \citet{DePasquale03} used the flux ratio of optical and X-ray afterglows, and \citet{Jakobsson04} used the spectral index ($\beta_{\rm OX} < 0.5$) obtained using optical to X-ray spectral energy distribution (SED) to classify these bursts. \citet{2009ApJ...699.1087V} suggested a more enlightened method and considered dark GRBs by $\beta_{\rm OX}$ (joint X-ray and optical spectral index) $<$ $\beta_{\rm X}$ (X-ray spectral index) -0.5 in the framework of the fireball model. 

There may be many potential causes for the optical darkness of GRB afterglows. 1) Dark GRB afterglows could be intrinsically faint \citep{Fynbo01}, for example, if the relativistic fireball is decelerated in a low-density surrounding medium \citep{Sari98}. 2) The high redshift origin \citep{Taylor98} of GRB afterglows, in such case the Ly-$\alpha$ forest will influence the optical wavelength \citep{Lamb00}. The cause of optical darkness due to high redshift origin is low and expected for $\sim$ 10-20 \% of dark bursts \citep{Perleya16}. 3) Obscuration scenario: This could be because of i) a high column density of gas or ii) dust in their host galaxies at larger distances so that the optical afterglow is very reddened \citep{Fynbo01}. The last scenario is particularly expected to be the most favoured cause for the optical darkness of GRB afterglows, as most of the host galaxies of dark bursts are detected at the optical wavelength.

The outline of the present work is as follows: In \S~\ref{multiwavlength observation and data analysis}, we present the multiwavelength observations and data reduction of \thisgrb. In \S~\ref{results for both the bursts}, we have given the results of this work. Finally, a discussion and a brief summary and conclusion of the work are given in \S~\ref{discussion} and \S~\ref{conclusiones}, respectively. The errors given in the paper are mentioned at 1 $\sigma$ level unless stated otherwise.

\section{Observations and Data Reduction}
\label{multiwavlength observation and data analysis}

\subsection{\bf High energy gamma-ray observations}

On 9$^{th}$ March 2015 at 22:59:50.66 UT (hereafter \fermiT), Gamma-Ray Burst Monitor (GBM, \citealt{2009ApJ...702..791M}) instrument of \fermi Gamma-ray space mission triggered a bright \thisgrb with a \tninty duration of about 52 s \citep{Roberts15}. The GBM light curve consists of two distinct emission episodes, the main episode followed by a soft, weak emission episode with a temporal gap of $\sim$ 200 s in between (see Fig. \ref{prompt_plot}). After $\sim$ 200 s (at 23:03:06 UT, \swiftT), the Burst Alert Telescope \citep[BAT;][]{bar05} instrument of \swift mission also triggered \thisgrb during the second weak emission episode \citep{Cummings15}. The mask-weighted light curve obtained using BAT observations also shows that \thisgrb consists of two separate episodes with a \tninty duration of 242 $\pm$ 6 s (15-350 \keV). The prompt emission of \thisgrb also was independently discovered by \kw mission \citep{Golenetskii15}. 

We retrieved photon events received by three sodium iodide (NaI) and one bismuth germanium oxide (BGO) detector (having the largest number of photon counts) of the GBM instrument onboard \fermi. These NaI and BGO detectors have the smallest boresight angle with respect to the GRB location. We carried out the temporal and spectral analysis of high energy $\gamma$-ray data as described in \citet{2023MNRAS.519.3201C}.

\subsection{\bf X-ray and Ultra-Violet observations}
The XRT instrument of \swift mission started follow-up observations of the BAT localisation region at 23:05:18.2 UT (131.5 s after the BAT trigger) to search for X-ray afterglow. An uncatalogued counterpart (X-ray) candidate was discovered at RA, DEC = 18h 28m 24.81, +86d 25' 43.6" (J2000) within the \swift BAT error circle with a 90\% uncertainty radius of 1.4 arcsec \footnote{\url{https://www.swift.ac.uk/xrt_positions/00634200/}} in the initial window timing (WT) mode exposure \citep{Cummings15}. The position of this fading afterglow was observed up to $\sim$ 4 $\times$ 10$^{5}$ s post-BAT detection. This work utilised X-ray afterglow data products, including both light curve and spectrum, obtained from the \swift online repository\footnote{\url{https://www.swift.ac.uk/}} hosted and managed by the University of Leicester \citep{eva07,eva09}. We conducted the analysis of the X-ray afterglow spectra acquired from the \swift XRT using the X-Ray Spectral Fitting Package (\sw{XSPEC}) package \citep{1996ASPC..101...17A}. The XRT spectra were analysed within the energy range of 0.3-10 \keV. For the XRT spectral analysis, we used an absorbed power-law model to explain the spectral properties of the X-ray afterglow of GRB 150309A. We considered both the photoelectric absorption from the Milky Way galaxy (\sw{phabs} model in \sw{XSPEC}) and the host galaxy of the GRB (\sw{zphabs} model in \sw{XSPEC})) along with the power-law model for the afterglow. The absorption due to our Galaxy was set as a fixed spectral parameter with a hydrogen column density of $\rm NH_{\rm Gal} = 9.05 \times 10^{20} \, \rm cm^{-2}$ \citep{2013MNRAS.431..394W}. However, the intrinsic absorption ({$\rm NH_{\rm z}$}) at the redshift value equal to two (the mean redshift $z$ value for a typical long-duration GRB) was permitted to vary freely. In addition, the photon index value of the power-law component was left to vary. We have used \sw{C-Stat} statistics for the spectral fitting of XRT data.

The Ultra-Violet and Optical Telescope \citep[UVOT;][]{2005SSRv..120...95R} onboard \swift began observing the XRT localisation region 140 s after the BAT trigger to search for UV/Optical afterglow. However, no credible Ultra-Violet (UV) or optical counterpart candidate was discovered within the \swift XRT error circle \citep{Cummings15, 2015GCN.17559....1O}. We extracted the UV afterglow data obtained using \swift mission following the method described in \citet{2021MNRAS.505.4086G}.

\subsection{\bf Optical and near-IR afterglow observations}

\begin{table*} 
\begin{center} 
\caption{The log of optical and near-infrared images obtained on the \thisgrb field (top panel for the afterglow and bottom panel for the potential host galaxy). The magnitude values in different filters are uncorrected for the reddening due to our galaxy. The non-detection upper limits values are listed at 3 $\sigma$. The magnitudes listed are given in the AB system. $^{*}$ marker denotes the magnitudes in the Vega system.}
\begin{tabular}{@{}lccccc@{}} 
\hline
Date   & T-\fermiT& Telescope/ & Filter/ & Exposure Time &  
Magnitude/ Upper limit \\ 
(UT, mid)  &  (s) & Instrument & Grism   &    (s)  & \\  
\hline 
09 Mar 2015, 23:05:53& 362  & 0.6m BOOTES-2/TELMA  & $--$ &       60   & $\geq$18.0   \\
09 Mar 2015, 23:05:53& 362  & 0.3m BOOTES-1        & $--$ &       60   & $\geq$16.5   \\
09 Mar 2015, 23:11:08& 677  & 0.6m BOOTES-2/TELMA  & $--$ &       120   & $\geq$20.2   \\
10 Mar 2015, 00:03:21& 3810  & 0.6m BOOTES-2/TELMA  & $i$ &       5400 &  $\geq$21.0   \\
10 Mar 2015, 01:22:34& 8563  & 0.6m BOOTES-2/TELMA  & $z$ &       3600 &  $\geq$19.2   \\
10 Mar 2015, 03:21:26& 15695  & 0.6m BOOTES-2/TELMA  & $Y$ &       9300 &  $\geq$18.0   \\
10 Mar 2015, 04:15&  18909 & GTC (CIRCE)    & $K_{\rm S}$ &      300   & 19.28 $\pm$ 0.11$^{*}$     \\ 
10 Mar 2015, 04:25& 19509  & GTC (CIRCE)    & $H$  &      300   & $\geq$ 21.4$^{*}$   \\
10 Mar 2015, 04:35& 20109  & GTC (CIRCE)    & $J$ &      300   &  $\geq$21.3$^{*}$   \\
10 Mar 2015, 05:09& 22149  & GTC (CIRCE)    & $K_{\rm S}$ &      300   &  19.50 $\pm$ 0.20$^{*}$   \\ 
\hline
03 Jul 2015, 00:30& 115.0625 (days)  & GTC (CIRCE)    & $K_{\rm S}$ &   1800   & $\geq$21.5$^{*}$   \\
03 Jul 2015, 00:40&  115.0696 (days) & GTC (CIRCE)    & $H$  &   1800   & $\geq$22.0$^{*}$   \\
\hline
Date   & T-\fermiT& Telescope/ & Filter/ & Exposure Time &  
Magnitude/ Upper limit \\ 
(UT, mid)  &  (days) & Instrument & Grism   &    (s)  & \\  
\hline 
19 Jul 2015, 01:31:50 & 131.1054  & GTC (OSIRIS)   & $r$  &   1800 (15x120 s)  & 25.26 $\pm$ 0.27 \\
24 Aug 2015, 00:34:18 & 167.0654  & GTC (OSIRIS)   & $i$  &   2160 (24x90 s)   &  24.89 $\pm$ 0.16 \\
25 Aug 2015, 23:33:37 & 169.0233  & GTC (OSIRIS)   & $i$  &   2160 (24x90 s)  &  25.09 $\pm$ 0.39 \\
07 Jul 2016, 01:32:06 & 485.1058 & GTC (OSIRIS)   &R1000B &  3600 (3x1200 s)  &   --    \\
30 Jul 2016, 00:26:21 &  508.0600 & GTC (OSIRIS)   & $g$  &   1200 (8x150 s)  &  25.56 $\pm$ 0.27 \\
30 Jul 2016, 01:20:48 & 508.0979  & GTC (OSIRIS)   & $i$  &   900 (10x90 s)  &  24.80 $\pm$ 0.22  \\
11 Mar 2021, 05:54:10 & 2193.2879  & GTC (OSIRIS)   & $z$  &   1890 (42x45 s) &   $\geq$24.4   \\
\hline
\label{optical/NIR observations} 
\end{tabular} 
\end{center} 
\end{table*} 
 
For \thisgrb, soon after the detection of X-ray counterpart by \swift XRT, many ground-based optical observatories (including BOOTES robotic telescope) started follow-up observations to search for the optical counterpart of \thisgrb, although no optical afterglow (from early to late phases) candidate consistent with the BAT position was detected. Our later optical non-detections are consistent with the optical limits given by \citep{2015GCN.17565....1R} and \citep{2015GCN.17570....1M}.

\begin{figure}[ht] 
\begin{center}
\includegraphics[angle=0,scale=0.4]{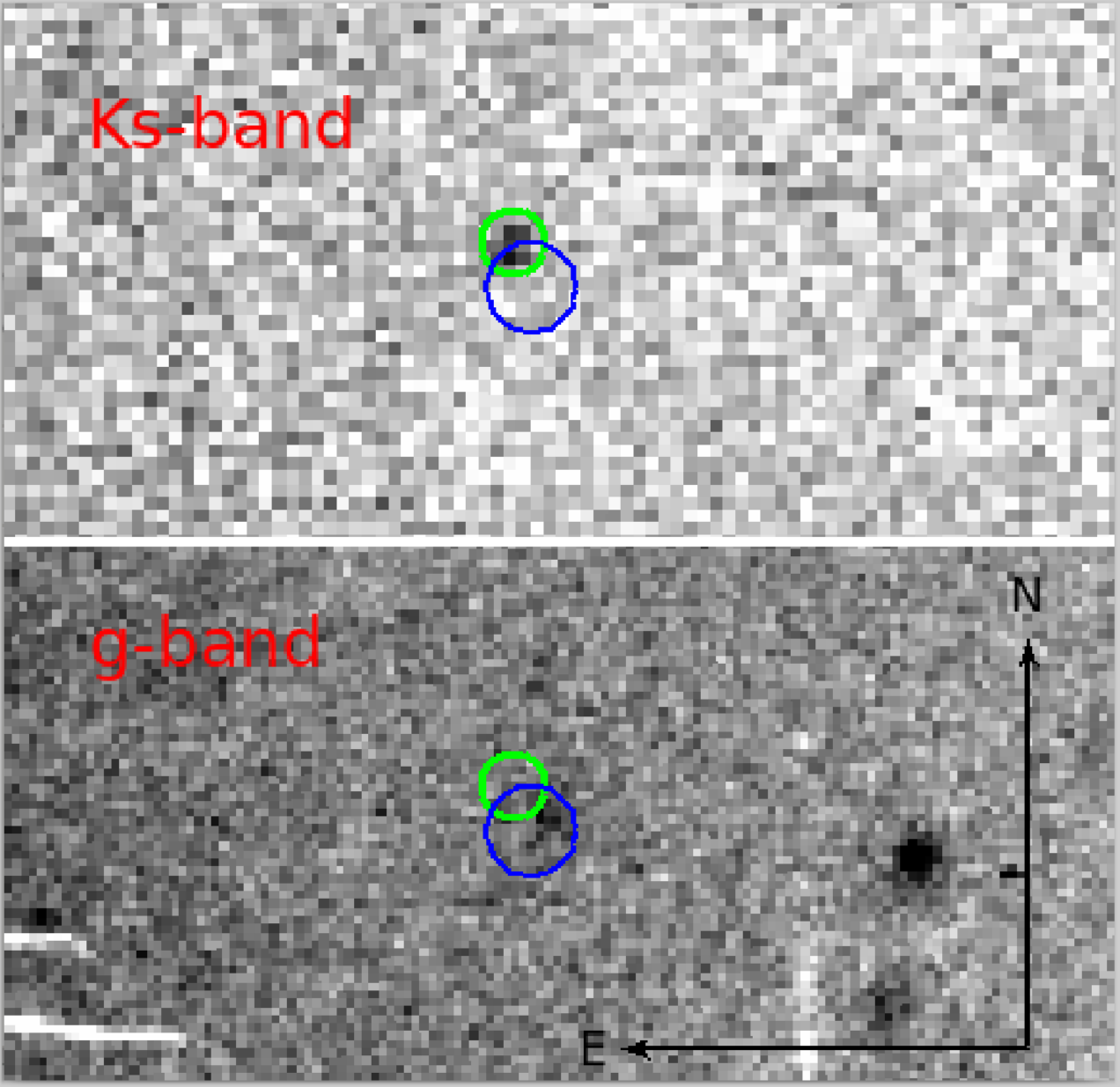} 
\caption{{\bf The field of \thisgrb.} {Top panel:} The $K_{\rm S}$-band discovery image (in green circle) of the afterglow of \thisgrb taken at the GTC (+CIRCE) in 2015. The near-IR afterglow is close to the centre of the image. {Bottom panel:} The late-time $g$-filter image of the field observed using the GTC in July 2016. The potential host galaxy is close to the centre of the image (within the XRT error circle of radius 1.4 arcsec, shown with a blue circle). North is shown in the upward direction, and East is shown towards the left direction.}
\label{carta JHK}
\end{center} 
\end{figure}

We triggered the Target of opportunity (ToO) follow-up observations in the near-IR (JHK) starting 5.0~hr after the prompt discovery with the GTC equipped with the CIRCE instrument at the Spanish Observatory of La Palma. We detected a potential NIR counterpart candidate using GTC, although we could not find any afterglow emission at optical wavelengths. Additional near-IR observations were made at GTC on July 3 2015 (i.e. 114 days post-burst). Table \ref{optical/NIR observations} displays the optical and NIR observing log of \thisgrb; the magnitude values listed in the table are given in the AB system. The $K_{\rm S}$-filter image of the \thisgrb field taken with the GTC (+CIRCE) in 2015 is shown in Fig. \ref{carta JHK}.

To determine the magnitudes from the optical and NIR frames, we utilised the DAOPHOT routine under Image Reduction and Analysis Facility (IRAF)\footnote{IRAF is distributed by the NOAO, National Optical Astronomy Observatories, which are operated by USRA, the Association of Universities for Research in  Astronomy, Inc., under cooperative agreement with the US National Science Foundation. NSF.}.

\subsection{\bf Search for potential Host Galaxy of \thisgrb}
\label{Host Galaxy}

We performed deep optical imaging in $griz$ filters using the GTC telescope (see Table \ref{optical/NIR observations}). The deeper late-time optical photometric observations were gathered on July 19, August 24, August 25 (2015), July 07, July 30 (2016), and March 11 (2021). We carried out image pre-processing, such as dark subtraction and flat-fielding, using IRAF routines. Further, we performed the photometry of cleaned GTC images using standard IRAF software. To calibrate the instrumental magnitude, we completed the field calibration using the Sloan Digital Sky Survey (SDSS) DR12 catalogue \citet{ala15}. AT the position RA= 18:28:24.67, and DEC= +86:25:44.16 (J2000), we detected a faint galaxy with the following magnitude values $g$ = 25.56 $\pm$ 0.27, $r$ = 25.26 $\pm$ 0.27, $i$ = 24.80 $\pm$ 0.22, and $z$ $\geq$ 24.4. The position of the candidate host galaxy is consistent with the XRT position (as shown in Fig. \ref{carta JHK}). We also performed spectroscopy observations (covering the 3700 -- 7750 \AA wave range) with an exposure time of 3x1200 s on July 07 2016, using GTC. We used standard procedures for analysing the Optical System for Imaging and low-Intermediate-Resolution Integrated Spectroscopy (OSIRIS) spectra. Table \ref{optical/NIR observations} shows the optical and NIR photometry of the potential host. The reported magnitude values are not corrected for Galactic reddening. The calculated Galactic reddening value is E(B-V) = 0.1206 \citep{sch11}. The potential host galaxy of \thisgrb is shown in Fig. \ref{carta JHK}. A detailed method of host galaxy data analysis of GTC data is described in \citet{2021MNRAS.505.4086G}.

\section{Results} 
\label{results for both the bursts} 

\subsection{\bf Prompt characterise and classification}
In the following subsections, we present the results of the prompt analysis of \thisgrb using high-energy observations.  

\subsubsection{\bf Light curve and spectral analysis:} 

\begin{figure}[ht]
\centering
\includegraphics[scale=0.31]{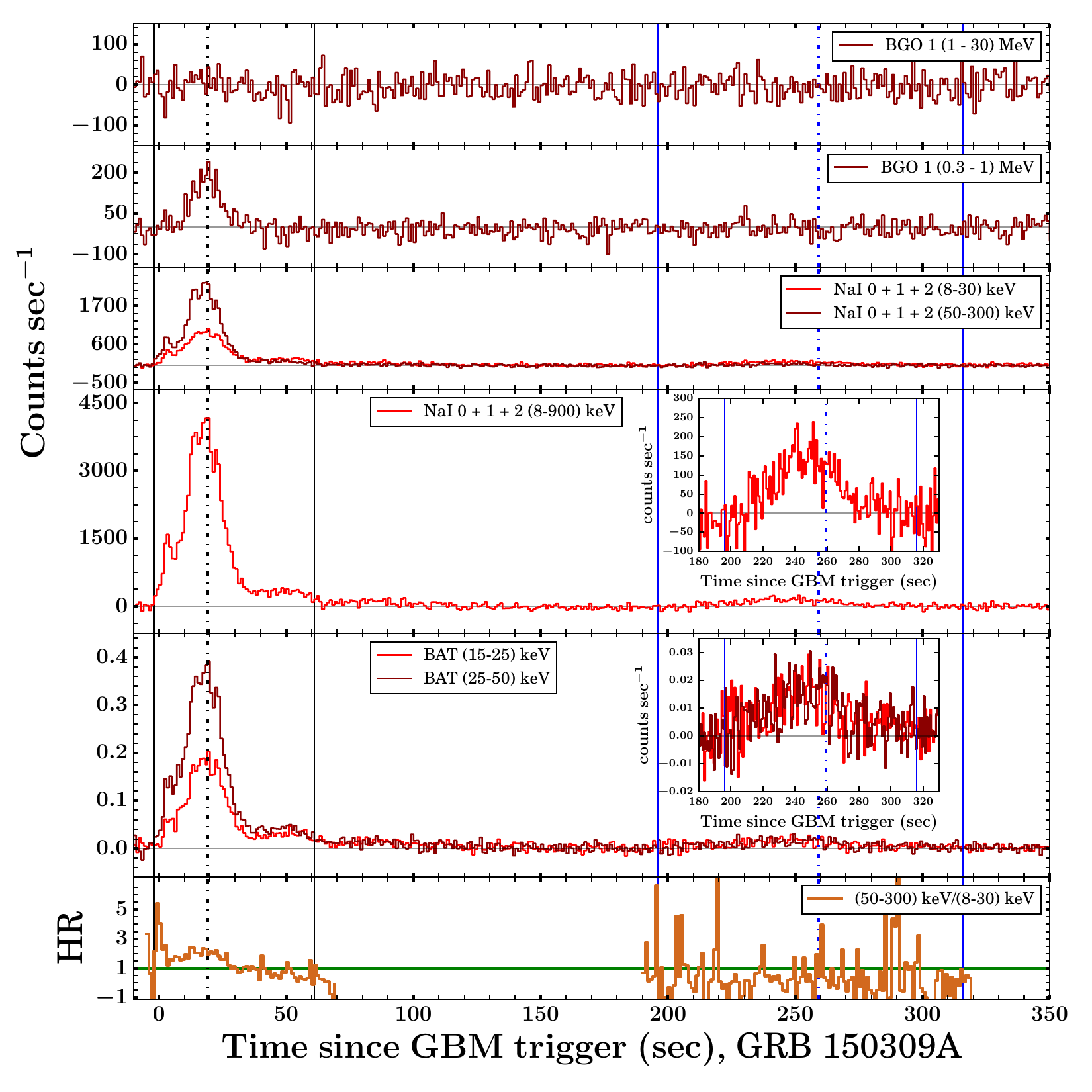}
\caption{{\bf Energy-resolved \fermi GBM and \swift BAT gamma-ray light curves of \thisgrb:} The multi-channel background-subtracted prompt emission light curve (top five panels) consists of two emission episodes with a significant temporal gap. The first and second episodes are shown with solid black and blue vertical lines. The peak times during both episodes are shown using black and blue vertical dashed-dotted lines, respectively. The insets in panels four and five show the zoomed version of the second episode based on GBM and BAT observations. The last panel shows the evolution of the hardness ratio during both episodes. The green horizontal solid line indicates the unit value of the hardness ratio.}
\label{prompt_plot}
\end{figure}

The energy-resolved $\gamma$-ray light curve for \thisgrb, along with the evolution of hardness ratio (HR), is presented in Fig. \ref{prompt_plot}. The light curves show two episodic emissions with a temporal gap in between. The first main episode is followed by a soft, weak episode with a very large ($\sim$ 200 s) temporal gap in between. Generally, two-episodic GRBs' prompt emission light curves consist of three different sub-classes. They follow a soft, weak emission episode preceding the main episode, the main emission component followed by a soft, weak emission component, or both the emission episodes have comparable intensity. The second sub-class (the main emission component followed by a soft, weak emission component) is very rare \citet{2018ApJ...862..155L} and \thisgrb belongs to this rare class. The evolution of HR shows that for \thisgrb, the first episode is harder than the subsequent episode, which is also seen from the very low signal for the second episode in the BGO 0.3-1 MeV light curve.

We performed the time-integrated and time-resolved spectral analysis using data from GBM detectors with the strongest signal: NaI 0, NaI 1, and NaI 2 (energy range 8-900 \keV). We also selected the BGO detector (energy range 250-40000 \keV) closest to the GRB direction (BGO 0). We generated the custom response matrices and spectra files using the publicly available \texttt{gtburst}\footnote{\url{https://fermi.gsfc.nasa.gov/ssc/data/analysis/scitools/gtburst.html}} software. We have used the Multi-Mission Maximum Likelihood framework \citep[\sw{3ML}\footnote{\url{https://threeml.readthedocs.io/en/latest/}}]{2015arXiv150708343V} software. A detailed spectral analysis method is described in \citet{2023MNRAS.519.3201C}. The time-integrated spectra of the first episode (\fermiT-1.35 to \fermiT+ 60.00 s) and second episode (\fermiT+215.72 to \fermiT+ 272.95 s) are best fitted by the \sw{Band} function with following model parameters: \Ep (peak energy) = $169.53_{-5.11}^{+5.08}$, $33.07_{-5.47}^{+5.30}$ \keV, $\alpha_{\rm pt}$ (low energy spectral index) = $-0.60_{-0.03}^{+0.03}$, $-0.59_{-0.38}^{+0.38}$ and $\beta_{\rm pt}$ (high energy spectral index) = $-2.86_{-0.20}^{+0.20}$, $-2.66_{-0.31}^{+0.29}$, respectively. The calculated values of \Ep also support the harder nature of the first episode.

\begin{figure}[ht]
\centering
\includegraphics[scale=0.34]{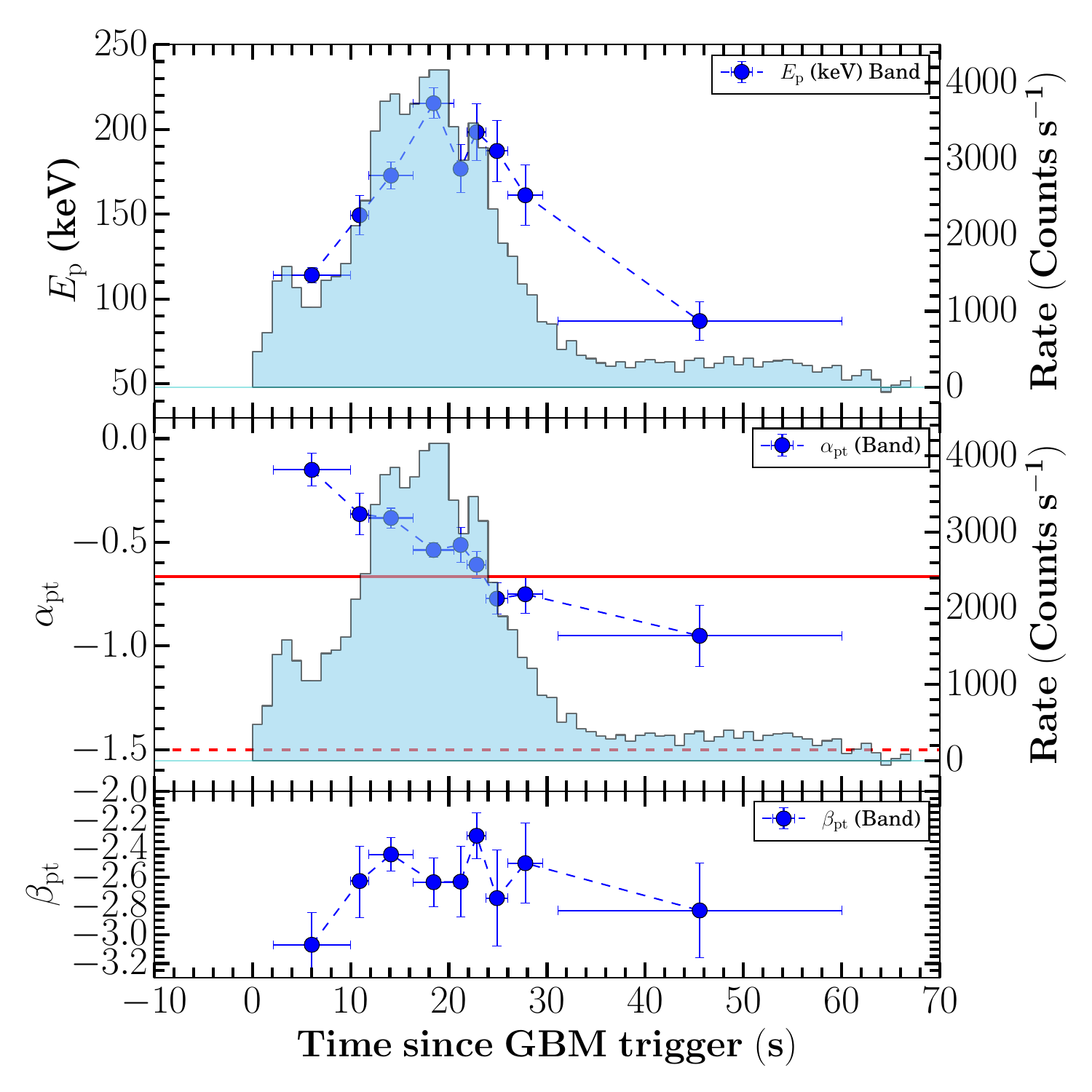}
\caption{{\bf Evolution of spectral parameters for \thisgrb.} (Top) The peak energy evolves with time and follows an intensity-tracking trend during the first episode. (Middle) The evolution of low energy spectral index $\alpha_{\rm pt}$ using \fermi GBM data. $\alpha_{\rm pt}$ seems to follow a hard to a soft trend during the first episode. The two horizontal lines are lines of death for synchrotron fast cooling ($\alpha_{\rm pt} = -3/2$, dotted dashed line) and synchrotron slow cooling ($\alpha_{\rm pt} = -2/3$, solid red line). (Bottom) The evolution high energy spectral index $\beta_{\rm pt}$ using \fermi GBM data.}
\label{TRS}
\end{figure}

Time-resolved spectroscopy of prompt emission of GRBs is a propitious method to explore the emission mechanisms (thermal or non-thermal) and to probe the correlations between spectral parameters of GRBs, which are yet an open question of GRB prompt emission physics. We divided the total emission duration of the first and second episodes of \thisgrb based on the Bayesian Block method, selected the energy channel from 8 - 900 \keV in the sodium iodide detector having a maximum count rate (NaI 1) as the Bayesian Block algorithm is believed to be the most acceptable method to recognise the intrinsic intensity variation in the prompt emission light curve of GRBs \citep{2013arXiv1304.2818S, 2014MNRAS.445.2589B}. We further consider only those Bayesian bins with statistical significance (S) $\geq$ 30 to ensure an excellent signal-to-noise ratio for the spectral analysis. These selection criteria provided the nine spectra for the first episodes; however, no bins for the second episode fulfilled our binning criteria. We used two models (\sw{Band}, and \sw{Cutoff-power law}) for time-resolved analysis. We notice that most of the time-resolved spectra (5/9) for the first episode are best fitted with the \sw{Band} function, and the remaining spectrum for the first episode is described with \sw{Cutoff-power law} model. The best-fit model parameters and associated errors are given in Table \ref{TRS_Table_Bayesian}. The evolution of spectral parameters along with the prompt emission light curve is shown in Fig. \ref{TRS}. As it can be seen from Fig. \ref{TRS} that \Ep is changing throughout the emission episodes, resulting in spectral evolution. The \Ep evolution seems to follow an intensity-tracking trend for the first episode. On the other hand, the evolution of $\alpha_{\rm pt}$ has a decreasing pattern with time. This signifies that the spectrum changes from harder to softer (a hard to soft trend), and the spectrum has a different origin than the synchrotron at the beginning (it exceeds the line of death for synchrotron slow cooling), but later synchrotron emission may dominate. 

We look for the correlations between spectral parameters obtained from time-resolved spectroscopy as they play an essential role in finding the emission mechanism of the prompt emission of GRBs. We look for the correlation between time-resolved \Ep - flux, $\alpha_{\rm pt}$-flux, and \Ep-$\alpha_{\rm pt}$ spectral parameters obtained using the \sw{Band} function based on GBM observations for each temporal bin of the first episode of \thisgrb. We found a strong correlation between the \Ep and the flux in 8 \keV to 30 MeV energy channels with a Pearson coefficient (r) and \sw{p-value} of 0.94 and 1.9 $\times$ $10^{-4}$. We found no correlation between $\alpha$ of \sw{Band} function and flux. Furthermore, we investigated the correlation between \Ep and $\alpha_{\rm pt}$. In this case, also, we did not find any correlation. Therefore, the first episodes of \thisgrb have the characteristics of an `intensity-tracking' GRB (the peak energy follows the intensity-tracking pattern) similar to many multi-episodic GRBs \citep{2020arXiv201203038L}. 

\subsubsection{\bf Classification of \thisgrb:} We calculated the prompt properties of \thisgrb (see Table \ref{tab:prompt_properties}) and compared them with a large population of GRBs to know the intrinsic class of \thisgrb. We mainly use four classifiers: spectral hardness-duration distribution, minimum variability timescale-duration distribution, spectral lag-luminosity correlation, and Amati correlation.  

\begin{figure*}
\centering
  \includegraphics[scale=0.29]{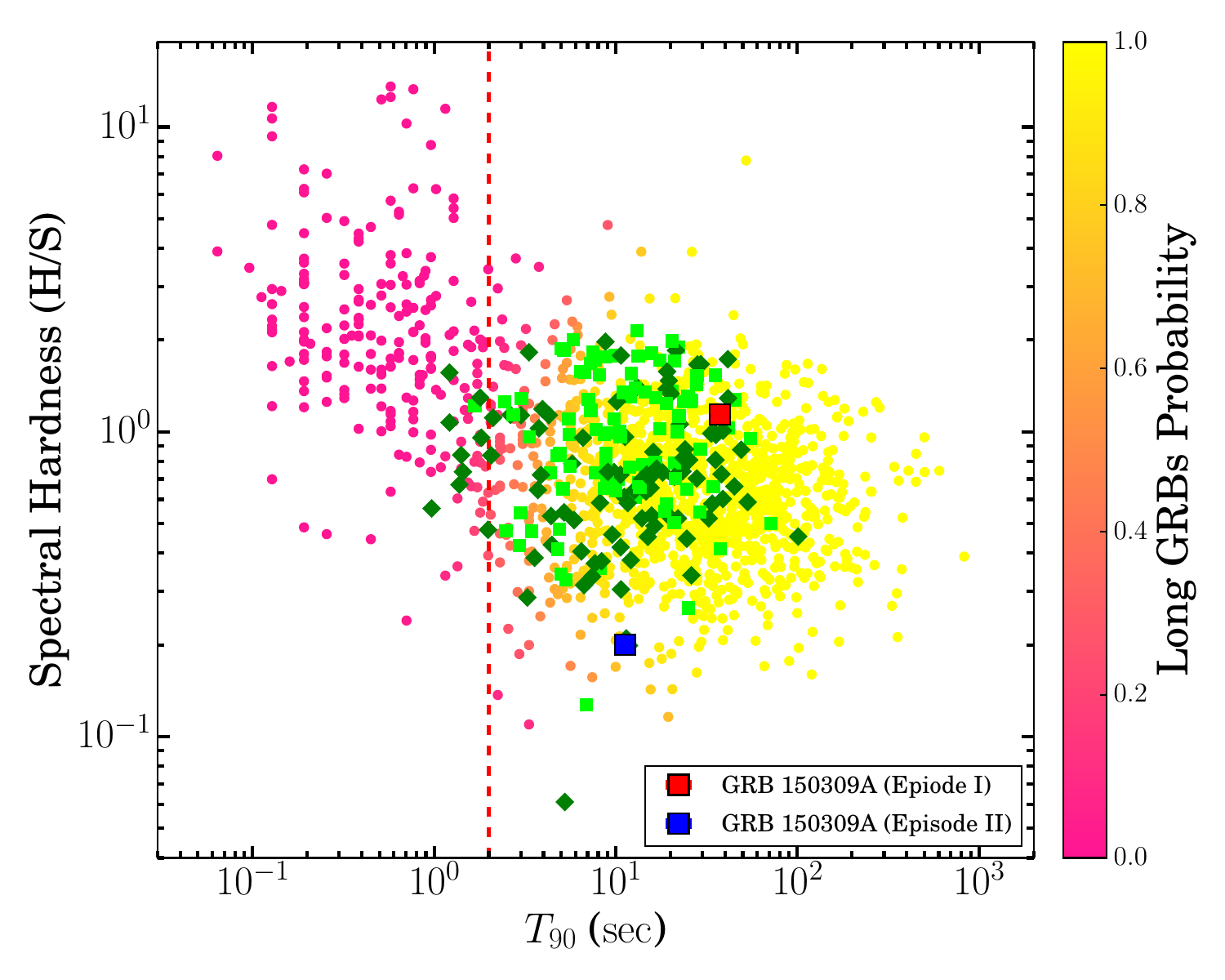}
  \includegraphics[scale=0.29]{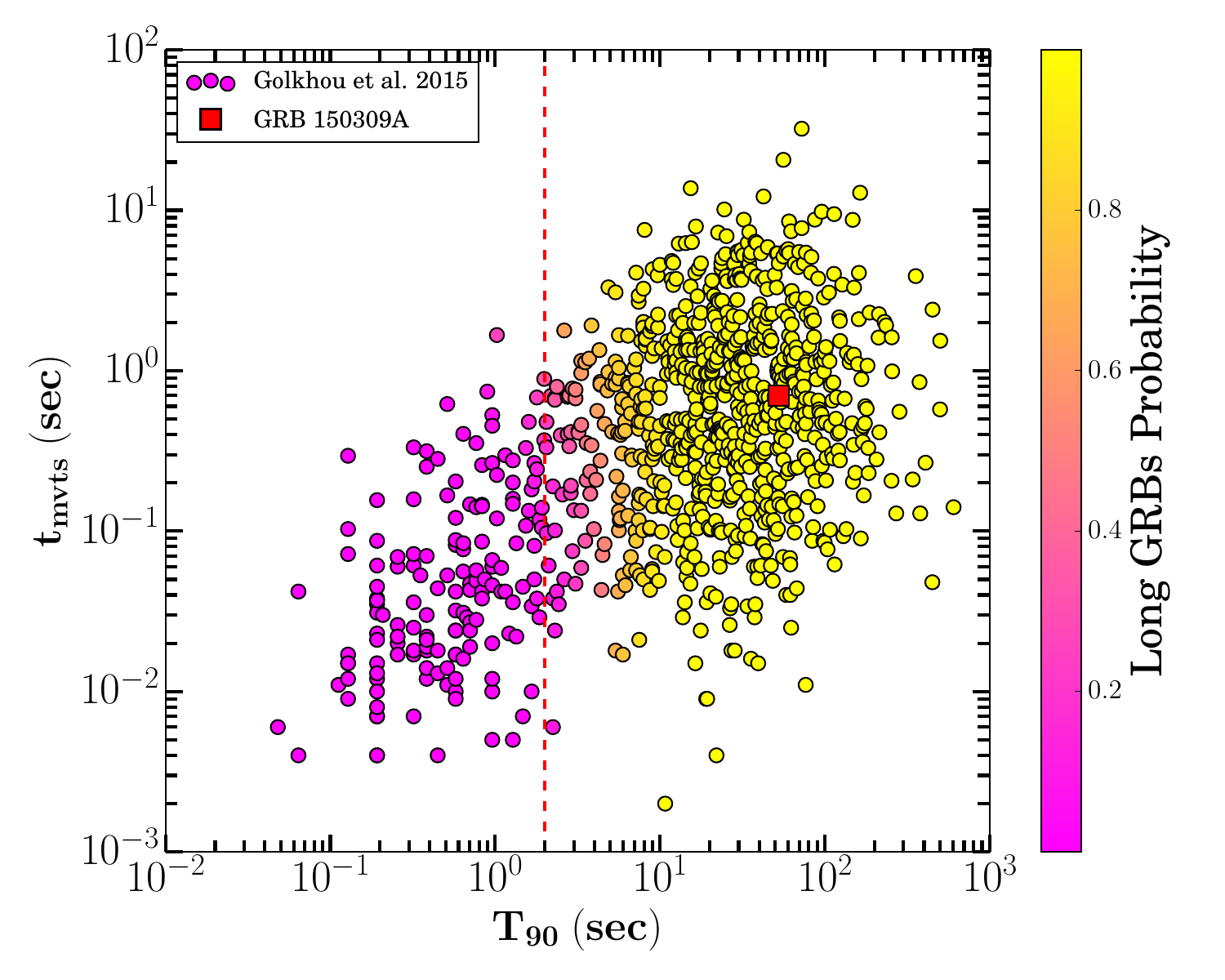}
  \includegraphics[scale=0.29]{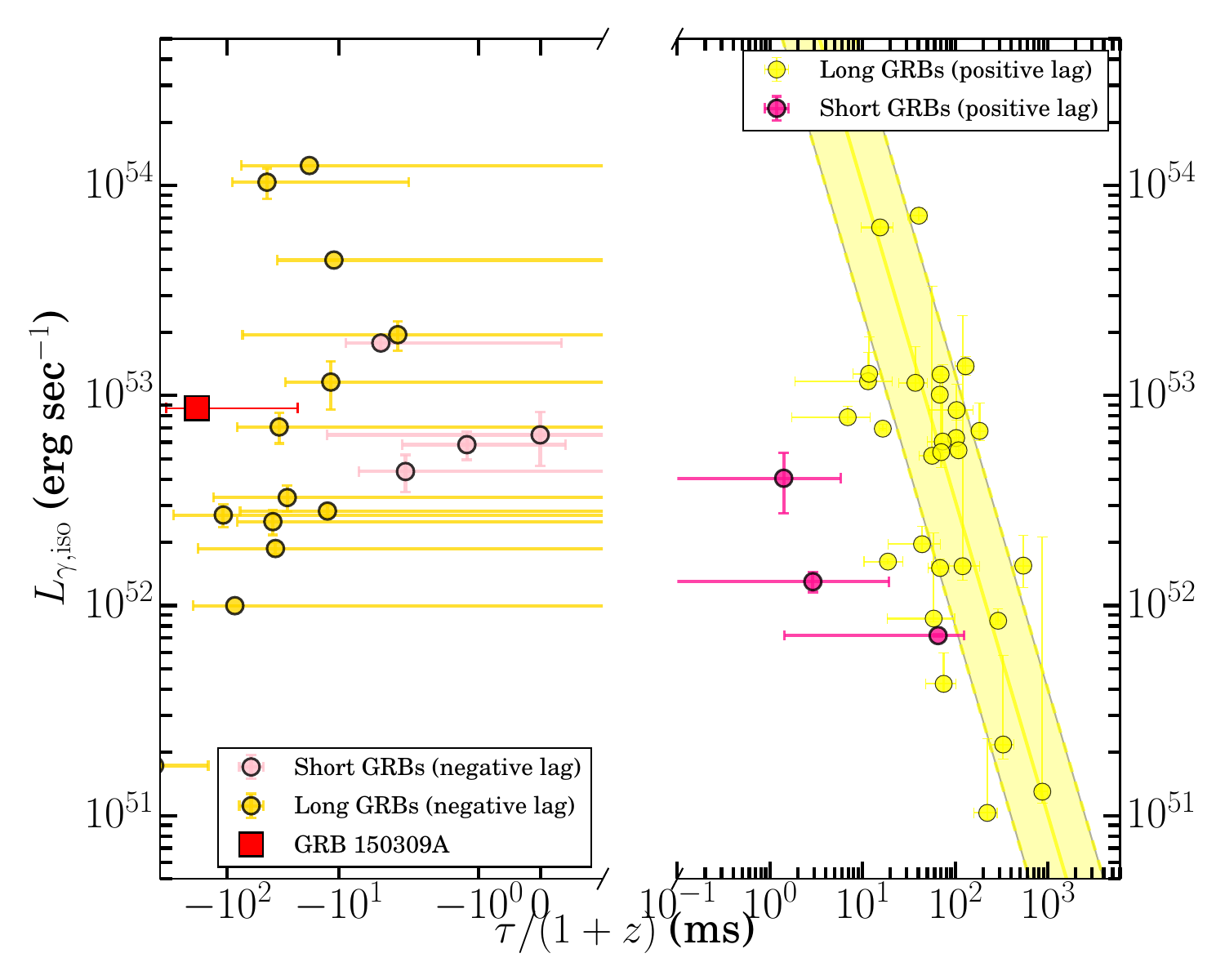}
\includegraphics[scale=0.29]{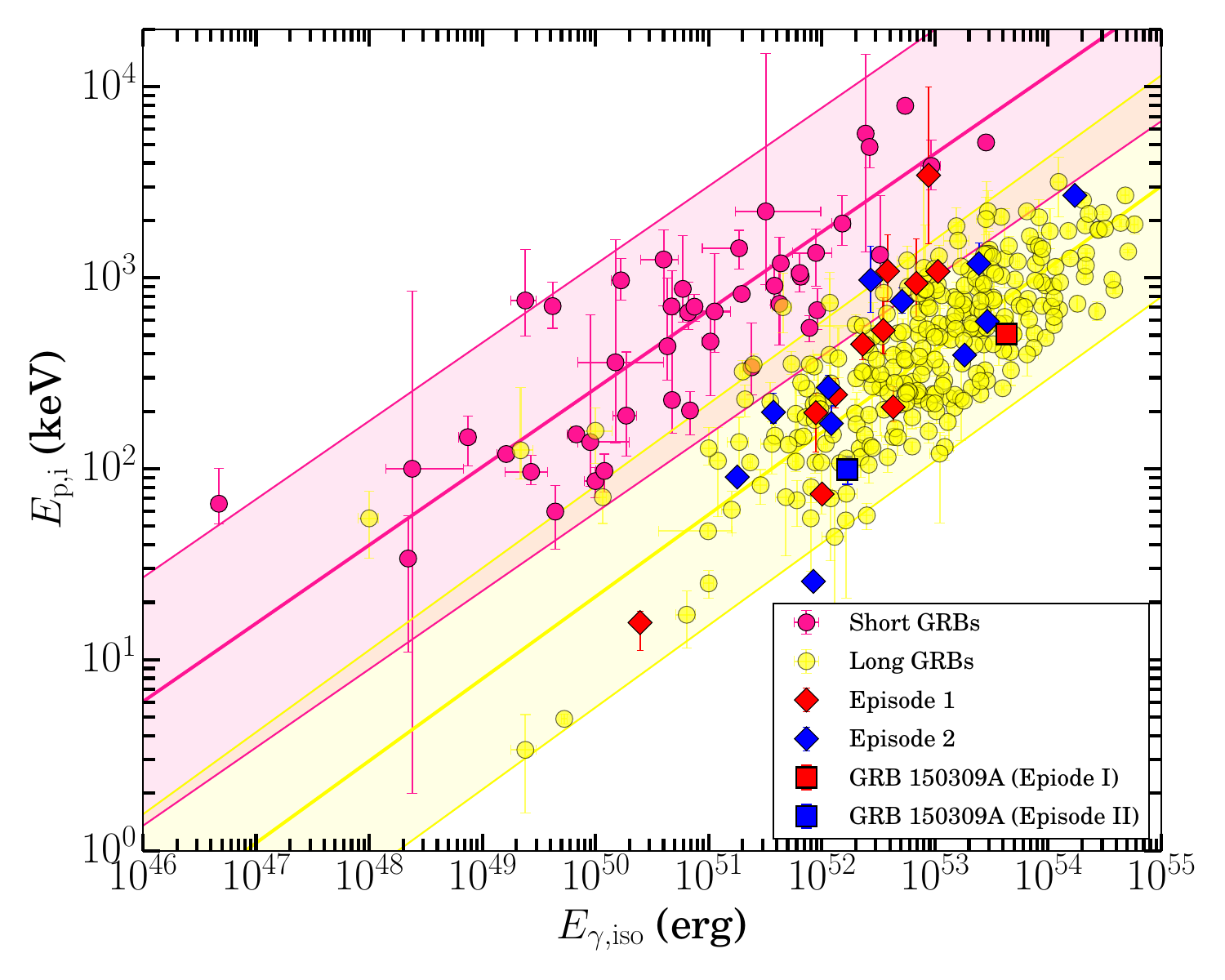}
\caption{ {\bf Classification of \thisgrb:} The conventional scheme of GRB classification based on the spectral hardness-duration distribution (top left), minimum variability timescale-duration distribution (top right), spectral lag-luminosity correlation (bottom left), and episode-wise Amati correlation (bottom right). The dashed lines show the boundary between long and short families of GRBs. The high-energy properties of \thisgrb (displayed with red/blue square symbol) are common among long soft bursts.}
\label{promptproperties}
\end{figure*}

To find the time-integrated HR of individual episodes, we divided the counts in soft (10 - 50 \keV) and hard (50-300 \keV) energy ranges using the three most brightest sodium iodide detectors of \thisgrb. We also calculated the \tninty duration of both the episodes and compared it with other two-episodic \fermi GRBs (with known redshift) studied by \citet{2020ApJ...898...42C}. The spectral hardness-duration distribution diagram (see Fig. \ref{promptproperties} (top left)) shows that both the episodes of \thisgrb have properties of long GRBs. The probabilities of a burst classified as a long burst are also shown in the background (obtained from \citealt{2017ApJ...848L..14G}).

The minimum variability time scales (\mvts) for long GRBs are typically longer than short GRBs. We calculated \mvts for \thisgrb utilising the Bayesian block method on the GBM light curve in 8-900 \keV. A more detailed method to determine \mvts is described in \citet{2018ApJ...864..163V}. The minimum variability timescale-duration distribution diagram for \thisgrb (red square) along with other GRBs sample \citet{2015ApJ...811...93G} is shown in Fig. \ref{promptproperties} (top right). The diagram shows that \thisgrb belongs to long GRBs. The probabilities of a burst classified as a long burst in the sample studied by \citet{2015ApJ...811...93G} are also shown.

Further, we calculated the spectral lag for \thisgrb following the method discussed in \citet{2023MNRAS.519.3201C}. We calculated the lag using prompt emission light curves (\fermiT-1.35 to \fermiT+60.00 s) in two energy ranges (15-25 \keV and 50-100 \keV). We obtained a negative lag for \thisgrb (possibly due to the superposition effect). The observed negative lag indicates that the low-energy photons were observed before the high-energy photons. We also calculated the isotropic peak luminosity of \thisgrb and plotted the spectral lag-luminosity correlation diagram (Fig. \ref{promptproperties} (bottom left)) for \thisgrb along with other samples studied by \citet{Ukwatta_2010}. We found that \thisgrb does not follow the spectral lag-luminosity anti-correlation.

According to the `Amati correlation' \citep{2006MNRAS.372..233A}, the time-integrated peak energies corrected to the rest frame ($E_{\rm p, i}$) are correlated with the prompt emission isotropic equivalent $\gamma$-ray energy ($E_{\gamma,\rm iso}$). This correlation is also studied and found valid for an episode-wise sample of GRBs \citet{2013MNRAS.436.3082B, 2020ApJ...898...42C}. Amati correlation is useful for classifying the GRBs. We studied the correlation for each episode of \thisgrb. To calculate $E_{\gamma,\rm iso}$ for both episodes, we calculated the time-integrated total energy fluence in 1-$10^{4}$ \keV energy channels in source rest frames. We have shown the episode-wise \thisgrb (red and blue squares) in the Amati correlation plane along with the other data points for long GRBs and short bursts studied in \citet{2020MNRAS.492.1919M} and \citet{2020ApJ...898...42C} (two-episodic GRBs). We noticed that both the episodes of \thisgrb are consistent with the Amati correlation of long bursts (see bottom right panel of Fig. \ref{promptproperties}).

\begin{table}[ht]
\caption{Prompt emission properties of individual episodes of \thisgrb. The episode-wise peak fluxes are measured in the 1-10,000 \keV energy range in the source frame. \mvts, HR, \Ep, peak flux ($F_{\rm p}$), $E_{\rm \gamma, iso}$, and $L_{\rm \gamma, iso}$ denote the minimum variability time scales, hardness ratio, peak energy, peak flux, isotropic equivalent $\gamma$-ray energy and isotropic equivalent $\gamma$-ray luminosity, respectively.}
\label{tab:prompt_properties}
\begin{center}
\begin{tabular}{|c|c|c|}
\hline
\bf {Prompt Emission Properties} & \bf {Episode 1 } & \bf {Episode 2} \\
\hline 
\tninty (s) & 37.43  & 11.27 \\ \hline
\mvts (s) & \multicolumn{2}{|c|} {$\sim$ 0.7} \\ \hline
HR  &   1.14 & 0.20 \\ \hline
\Ep (\keV)  &$169.53_{-5.11}^{+5.08}$  & $33.07_{-5.47}^{+5.30}$ \\ \hline
$F_{\rm p}$  ($\rm 10^{-7} ~erg ~cm^{2} ~s^{-1}$) & 7.11 & 0.27 \\\hline
$E_{\rm \gamma, iso}$ (erg) & $43.09 \times 10^{52}$ & $1.68 \times 10^{52}$\\\hline
$L_{\rm \gamma, iso}$ (erg $\rm s^{-1}$) & 8.72$ \times 10^{52}$ &  -\\ \hline
Redshift $z$ & \multicolumn{2}{|c|} {2.0}  \\ \hline
Spectral lag (s) &  {-0.56$^{+0.49}_{-0.50}$} & -  \\
\hline
\end{tabular}
\end{center}
\end{table}

\subsection{\bf Afterglow Characteristics}
In the following subsections, we present the results of the afterglow analysis of \thisgrb using X-ray to NIR band observations.  

\subsubsection{\bf The nature of X-ray afterglow of \thisgrb} 
\label{x-ray}         

The X-ray afterglow observations using \swift/XRT shows a very steep decline $\rm \alpha_{\rm x1}$ = $-2.13^{+0.09}_{-0.09}$ from \swiftT+ 123 s to \swiftT+397 s followed by a shallower decline rate of $-1.05^{+0.03}_{-0.03}$ after \swiftT+397 s (see Fig. A1 of the appendix). 
Further, we carried out \swift XRT spectral analysis for the temporal bins selected before and after the temporal break following \citep{2021MNRAS.505.4086G}. We calculated the spectral parameters for all the Photon Counting (PC) mode observations (interval after the temporal break) as the photon index value of $\Gamma_{\rm XRT} = 1.78^{+0.10}_{-0.10}$, and the intrinsic hydrogen density column $N_{\rm H, z} = 2.43^{+0.62}_{-0.57} \times 10^{22}$~cm$^{-2}$, considering $z$ = 2. Our intrinsic hydrogen density column measurement shows clear evidence of excess absorption over Galactic hydrogen column density. For the interval before the break, we have frozen the  $N_{\rm H, z}$ from the PC mode data and calculated $\Gamma_{\rm XRT} = 1.90^{+0.05}_{-0.05}$. Further, we also checked the evolution of photon indices from the \swift burst analyser page \footnote{\url{https://www.swift.ac.uk/burst_analyser/00634200/}} and noted that photon indices do not change significantly during the entire emission phase. The unavailability of obvious spectral evolution between the data before and after the break plus the large change in the temporal index rules out the possibility of relating the temporal break at \swiftT+397 $\pm$ 48.60 s to the crossing of the cooling break frequency ($\nu_{\rm c}$) across the X-ray wavelength.

The early steep decay emission observed before the break time ($<$ 400 s) could be interpreted as the tail (high-latitude emission) of the prompt emission. In this case, if the emission is purely non-thermal and consists of a single power-law, the temporal index can be expressed as $\alpha $\,=2\,+\,$\beta$ \citep{2000ApJ...541L..51K}. For the present burst, $\beta_{1}$ (X-ray spectral index during step decay phase)\,=0.90$\pm 0.05$, which results in an expected temporal index of $\sim$\,2.90, inconsistent with the observed decay (2.13\,$\pm 0.09$). Although, $\alpha $-$\beta$ relation during the early steep decay phase can be explained considering that the peak energy (\Ep) and flux during this phase rely on the viewing angle \citet{2011A&A...529A.142R}. This multi-peaked behaviour observed during the prompt emission can also cause the temporal indices inconsistent with the simple correlation between $\alpha $-$\beta$ during the high-latitude emission \citep{2009MNRAS.399.1328G}.

At later times ($>$400 s), the observed temporal and spectral indices ($\alpha_{2}$\,=\,1.05\,$\pm$\,0.03, $\beta_{2}$\,=0.78$\pm$\,0.10) are dominated by non-thermal emission because of the interaction of ejecta with the ambient medium (see section \ref{sed}). To investigate the nature of the circumburst medium (ISM or wind), we apply the closure relations between the temporal and spectral index during the shallow decay phase.

1.- Case when $\nu_{\rm X}$\,$<$\,$\nu_{\rm c}$:
\begin{itemize}
\setlength{\itemindent}{.5in}
        \item $\alpha_{\rm ISM}$\,=\,3$\beta_{2}$/2\,= 1.17$\pm0.15$
        \item $\alpha_{\rm wind}$\,=\,(3$\beta_{2}$\,+\,1)/2\,=\,1.67$\pm0.15$
\end{itemize}

2.- Case when $\nu_{\rm X}$\,$>$\,$\nu_{\rm c}$ (both mediums are indistinguishable):
\begin{itemize}
\setlength{\itemindent}{.5in}
        \item $\alpha$\,=\,(3$\beta_{2}$\,-\,1)/2\,=\,0.67$\pm0.15$
\end{itemize}

The case for an ISM environment with $\nu_{\rm X}$\,$<$\,$\nu_{\rm c}$ seems to be the best fit for our data. All these relations assume an electron spectral index, $p$\,$>$\,2. The electron spectral index obtained through the closure relations is $p$\,= 2.56 $\pm0.20$. Based on closure relations, all the cases with the wind ($\rho$ $\propto$ $k^{-2}$) environment seem to be completely ruled out. Further, using the calculated values of temporal ($\alpha_{2}$\,=\,1.05\,$\pm$\,0.03) and spectral ($\beta_{2}$\,=0.78$\pm$\,0.10) indices during $\nu_{\rm X}$\,$<$\,$\nu_{\rm c}$ spectral regime and equation 1 of \citet{2014A&A...567A..84M}, we computed the density profile ($k$ = 0 for ISM and $k$ = 2 for wind) of the environment of \thisgrb. We obtained $k$ $\sim$ -1.3, suggesting an intermediate density between ISM and wind ambient medium.  

Additionally, we computed the value of jet kinetic energy ($E_{\rm K, iso}$) for an ISM environment with $\nu_{\rm X}$\,$<$\,$\nu_{\rm c}$ spectral regime utilising the equation 11 of \citet{2018ApJS..236...26L}. We assumed typical micro-physical parameters (jet is due to Wiebel shocks and $\epsilon_{e}$ $\approx$ $\sqrt {\epsilon_{B}}$) to those found in the case of other well-studied GRBs, such as $\epsilon_{B}$ (energy fraction in the magnetic field) = 0.01, $\epsilon_{e}$ (energy fraction in electrons) = 0.1, and number density $n_{0}$ = 1 \citep{2002ApJ...571..779P}. We obtained $E_{\rm K, iso}$ = 5.36 $\times ~10^{53}$ erg for \thisgrb.

\subsubsection{\bf A potential near-IR afterglow of \thisgrb and its confirmation} 
\label{sin emision} 

The fact that the XRT discovered the X-ray counterpart of \thisgrb \citep{Cummings15} enabled us to quickly identify a potential near-IR afterglow within the XRT X-ray error box, at coordinates RA (J2000) = 18 28 25.00, Dec(J2000) = +86 25 45.10 ($\pm$ 0$^{\prime\prime}$.4). The source remained stable (within errors) in brightness during the CIRCE observation window, with $K_{\rm S}$ = 19.28 $\pm$ 0.11 (Vega) and $K_{\rm S}$ = 19.50 $\pm$ 0.20 (Vega) at the beginning and end of the observation. The object was very red (Fig. \ref{carta JHK}), and only upper limits were derived at the bluest NIR bands: $J$ $\geq$ 21.3 (Vega) and $H$ $\geq$ 21.4 (Vega). We calibrated the NIR observations utilising the Two Micron All Sky Survey (2MASS) catalogue.

In order to confirm whether the highly reddened object (H-$K_{\rm S}$) $\geq$ 2 is the afterglow or an extremely reddened host galaxy, a second epoch observation at GTC (+CIRCE) was conducted on July 3. No source was detected with the following upper limits: $H$ $\geq$ 22.0 (Vega) and $K_{\rm S}$ $\geq$ 21.5 (Vega), implying that the afterglow faded by more than 2 mag in the $K_{\rm S}$-band after 114 days. This fading behaviour confirms that the highly reddened object is the NIR afterglow of \thisgrb. 

\subsection{\bf Spectral Energy Distribution of \thisgrb} 
\label{sed}         

\begin{figure}
\begin{center} 
\includegraphics[angle=0,scale=0.35]{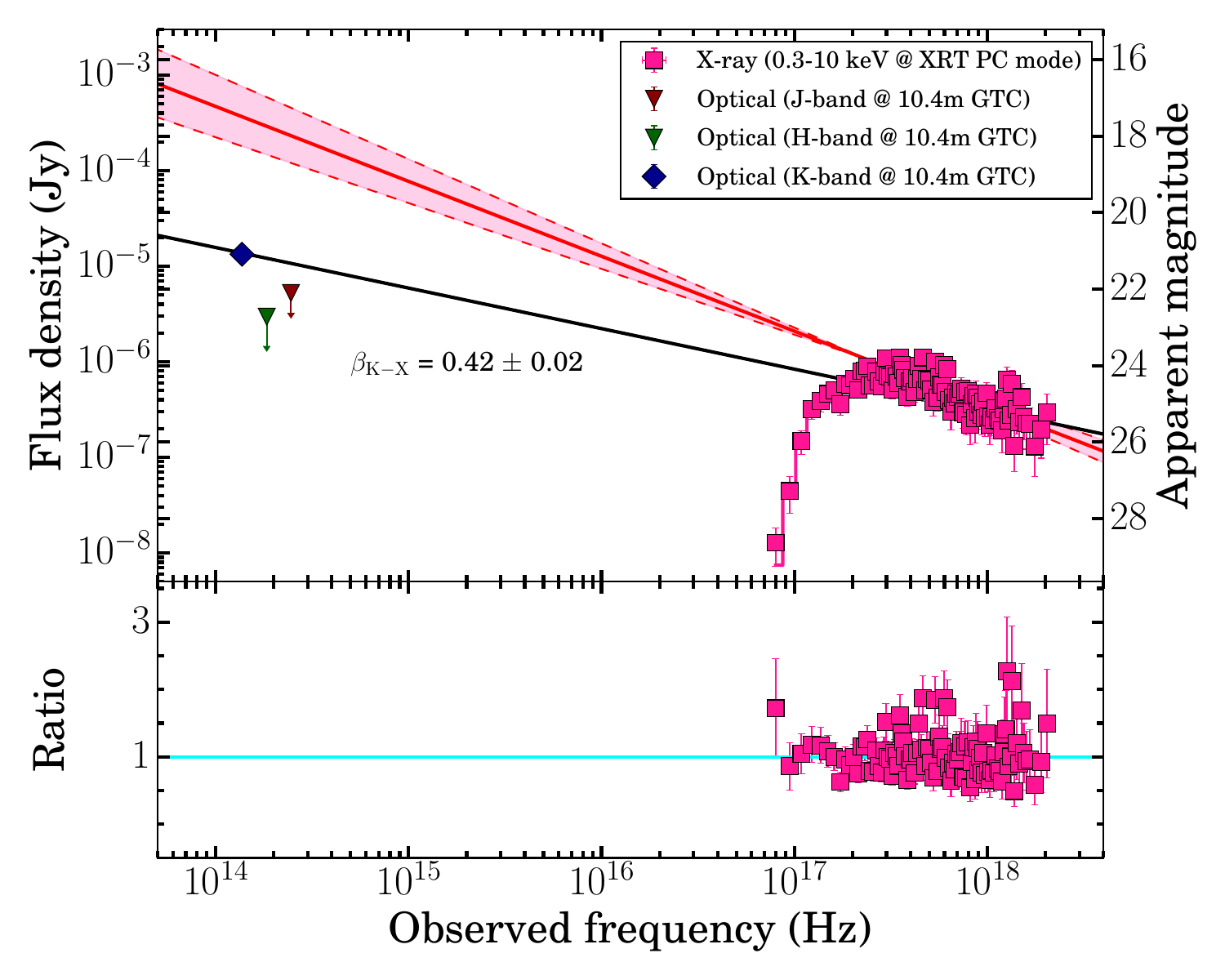} 
\caption{{\bf The SED of \thisgrb afterglow.} {Top panel:} The SED at \fermiT + 5.2 hours using the simultaneous X-ray data as well as the near-IR upper limits (and the K-band detection) reported in this paper. The solid red line indicates the X-ray spectral slop obtained using the best fit time-averaged PC mode spectrum, and the pink shaded region within the red dashed lines indicates the associated uncertainty with the X-ray spectral slop. The solid black line indicates the joint NIR-X-ray spectral index ($\beta_{\rm K-X}$). The left and right sides of the Y-axis, representing the flux density and magnitude values, are in the AB system. {Bottom panel:} Ratio of X-ray observed data and best-fit model. The horizontal cyan line indicates the ratio equal to one.}  
\label{OA sed}
\end{center} 
\end{figure}

SED analysis using multiwavelength data is very useful in constraining the location of synchrotron break frequencies and host extinction. We created the SED at \fermiT + 5.2 hours at the time of the GTC (+CIRCE) observations (to maximise the near-simultaneous broadband coverage). We have shown the SED for \thisgrb in Fig. \ref{OA sed}. The closure relations for the X-ray afterglow support that the cooling break frequency is beyond the X-ray spectral coverage ($>$ 10 \keV) for an ISM-like ($\rho$ = constant) ambient medium and slow-cooling regime at the epoch of the SED, supporting that no break in SED between X-ray and optical/NIR energies. Therefore, we extrapolated the unabsorbed spectral index of X-ray afterglow towards NIR wavelengths to calculate the extinction in the host galaxy. We noted that $K_{\rm S}$ band Galactic extinction corrected detection and upper limits in J and H filters using the GTC telescope are situated much below the extrapolated X-ray spectral slope (see Fig. \ref{OA sed}). Our SED analysis suggests that the host galaxy of the burst is highly extinguished, and \thisgrb is a dark burst. We determined the $K_{\rm S}$ band host extinction (A$_{K_{\rm S}}$) = 3.60$^{+0.80}_{-0.76}$ mag) utilizing the X-ray to NIR SED. The calculated host extinction in the $K_{\rm S}$ filter corresponds to A$_{\rm V}$  = 34.67$^{+7.70}_{-7.32}$ mag, suggesting that \thisgrb is one of the most extinguished bursts to date. We constrain the host galaxy extinction considering the Milky Way extinction law. We noted that such a high value of optical extinction (A$_{\rm V}$ =  34.67$^{+7.70}_{-7.32}$ mag) had been previously observed in the case of a few dark bursts, for example, GRB 051022 \citep{2007A&A...475..101C}, GRB 070306 \citep{2008ApJ...681..453J}, GRB 080325 \citep{2010ApJ...719..378H}, GRB 100614A, GRB 100615A \citep{2011A&A...532A..48D}.

\section{Discussion}
\label{discussion}

\subsection{\bf \thisgrb: A dark burst}

\begin{figure}[ht]
\centering
\includegraphics[angle=0,scale=0.35]{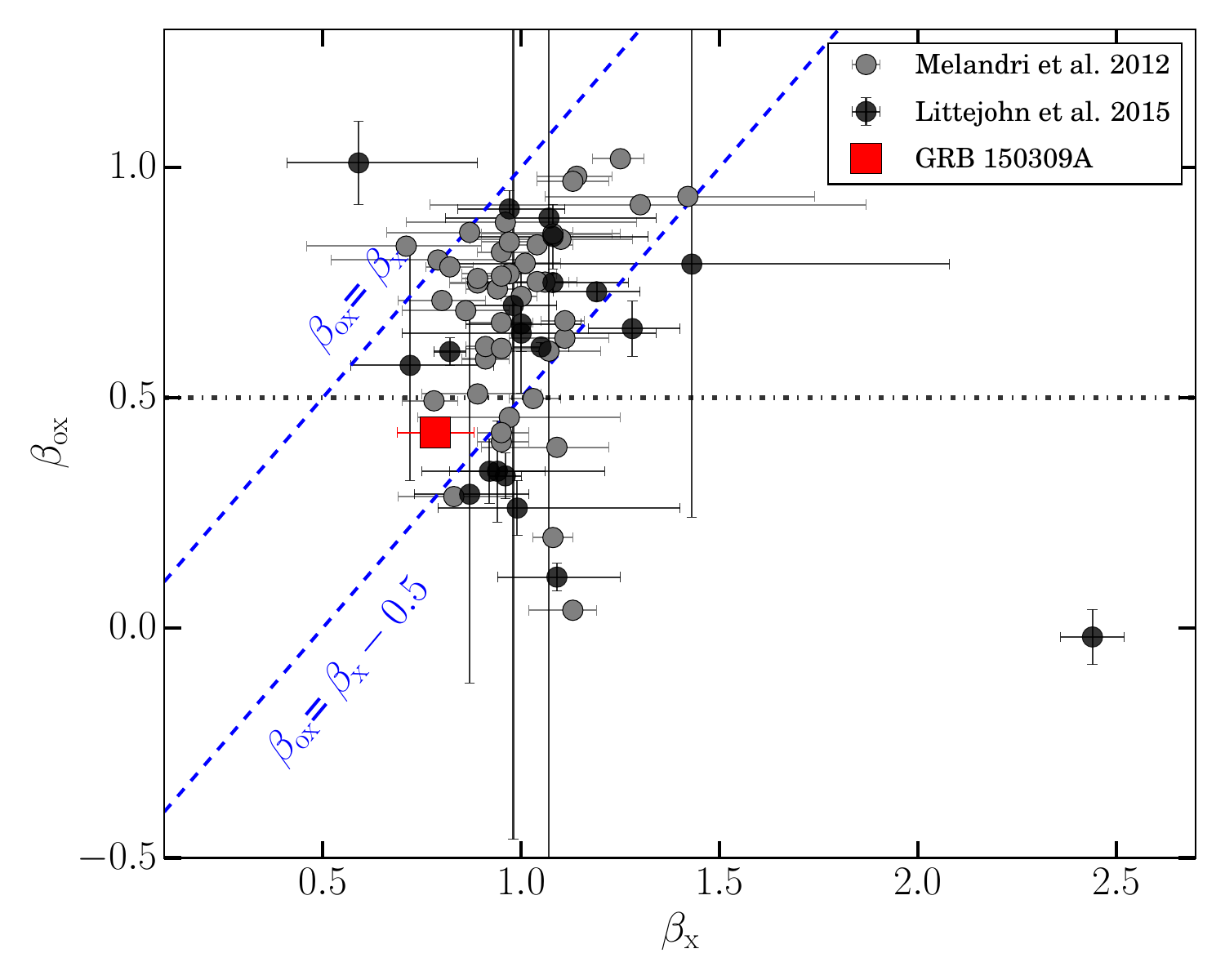}
\caption{The evolution of NIR-X-ray spectral index as a function of X-ray spectral index for \thisgrb (presented with a red square). We have also shown the $\beta_{\rm OX}$-$\beta_{\rm X}$ data points for other well-studied samples of \swift GRBs \citet{2012MNRAS.421.1265M, 2015MNRAS.449.2919L} for the comparison. The horizontal black dashed line shows the $\beta_{\rm OX}$ $=$ 0.5.}
\label{betaox}
\end{figure}

There are many possible explanations proposed for GRBs to be dark. In the case of \thisgrb, the optical non-detections up to deeper limits indicate dark behaviour following the definition proposed at the early stages of their discovery \citep{1998ApJ...493L..27G}. In addition, we calculated the joint NIR-X-ray spectral index ($\beta_{\rm K-X}$) for \thisgrb (the closure relation support for no spectral break between NIR to X-ray and cooling frequency lies beyond the X-ray energies) using the spectral energy distribution at \swiftT+5.2 hour and noted $\beta_{\rm K-X}$ = 0.42 $\pm$ 0.02 (see Fig. \ref{sed}). However, we could not estimate the optical-to-X-ray spectral index ($\beta_{\rm OX}$) at 11 hours post-trigger due to the unavailability of the optical observations near this epoch for \thisgrb. The calculated value of $\beta_{\rm K-X}$ also supports that \thisgrb is a dark burst such as GRB 051022 \citep{2007A&A...475..101C} and follows the definition described in \citet{Jakobsson04}. Further, we plotted the evolution of NIR-X-ray spectral index vs X-ray spectral index for \thisgrb (see Fig. \ref{betaox}). We have also shown the $\beta_{\rm OX}$-$\beta_{\rm X}$ data points for other well-studied samples of \swift GRBs \citet{2012MNRAS.421.1265M, 2015MNRAS.449.2919L} for the comparison. Furthermore, \citet{2009ApJ...699.1087V} suggested the possible range of joint optical-X-ray spectral slope in different possible regimes: $\beta_{\rm X}$ -0.5 $\leq$ $\beta_{\rm OX}$ $\leq$ $\beta_{\rm X}$ following the external forward shock synchrotron model. We found \thisgrb well satisfy the definition of darkness suggested by \citet{2009ApJ...699.1087V} in  $\nu < \nu_{\rm c}$ spectral regime.

\subsection{\bf Origin of the optical darkness of \thisgrb}

We detect a source at the edge of the \swift XRT error circle (1.4 arcsec) with $K_{\rm S}$ = 19.28 $\pm$ 0.11 mag (Vega) in the first set of $K_{\rm S}$-band observations. We do not detect the afterglow either in the $H$ or the $J$ band data. We detected it again at $K_{\rm S}$ = 19.50 $\pm$ 0.20 mag (Vega) in the second set of $K_{\rm S}$ data (the magnitude difference is $\leq$1-sigma, so no apparent fading on this timescale of 20 minutes). The position offset of the $K_{\rm S}$-band afterglow is 1.1-arcsec $\pm$ 1.5-arcsec (including 1.4-arcsec error circle from \swift/XRT, and about 0.25-arcsec from CIRCE/2MASS astrometry), so this is definitely in the positional uncertainty region. The clear evidence of the fading nature of the NIR source and its position consistent with the X-rays afterglow suggest that it is certainly the counterpart of the burst. Perhaps most interesting is the non-detection in H (bracketed in time by detection in the $K_{\rm S}$-band). The 2-sigma upper limit in H band observations is H $\geq$ 21.4 mag (Vega). This means that the counterpart is very red, with H-K $\geq$ 2.1 mag (95 \% confidence). Only two scenarios are possible: i) either a very dusty host galaxy with deeply embedded GRB (A$_V$ = 34.67$^{+7.70}_{-7.32}$ mag) or a very high-redshift Ly-alpha dropout ($z$ $\geq$10).

\subsubsection{\bf Dust extinguished scenario} 
\label{obscuration} 

The observed characteristic of \thisgrb indicates that it is positioned undoubtedly in the dark burst region of the $\beta_{\rm OX}$ $-$ $\beta_{\rm X}$ diagram (see Fig. \ref{betaox}). It is clear that the optical afterglow of \thisgrb could not be detected because of the obstruction of sight.
We carried out SED analysis (at \swiftT + 5.2 hours) to determine the host extinction at NIR wavelengths. Our SED analysis suggests $K_{\rm S}$ band brightness equal to $\sim$ 17.49 mag (the intrinsic afterglow brightness without any extinction obtained from the extrapolation of the unabsorbed spectral index of X-ray afterglow towards $K_{\rm S}$ band), from which we found a large host extinction at NIR wavelengths, A$_{K_{\rm S}}$ = 3.60$^{+0.80}_{-0.76}$ mag (see section \ref{sed}). 

\subsubsection{\bf A high redshift scenario} 

\begin{figure}[ht]
\centering
\includegraphics[angle=0,scale=0.35]{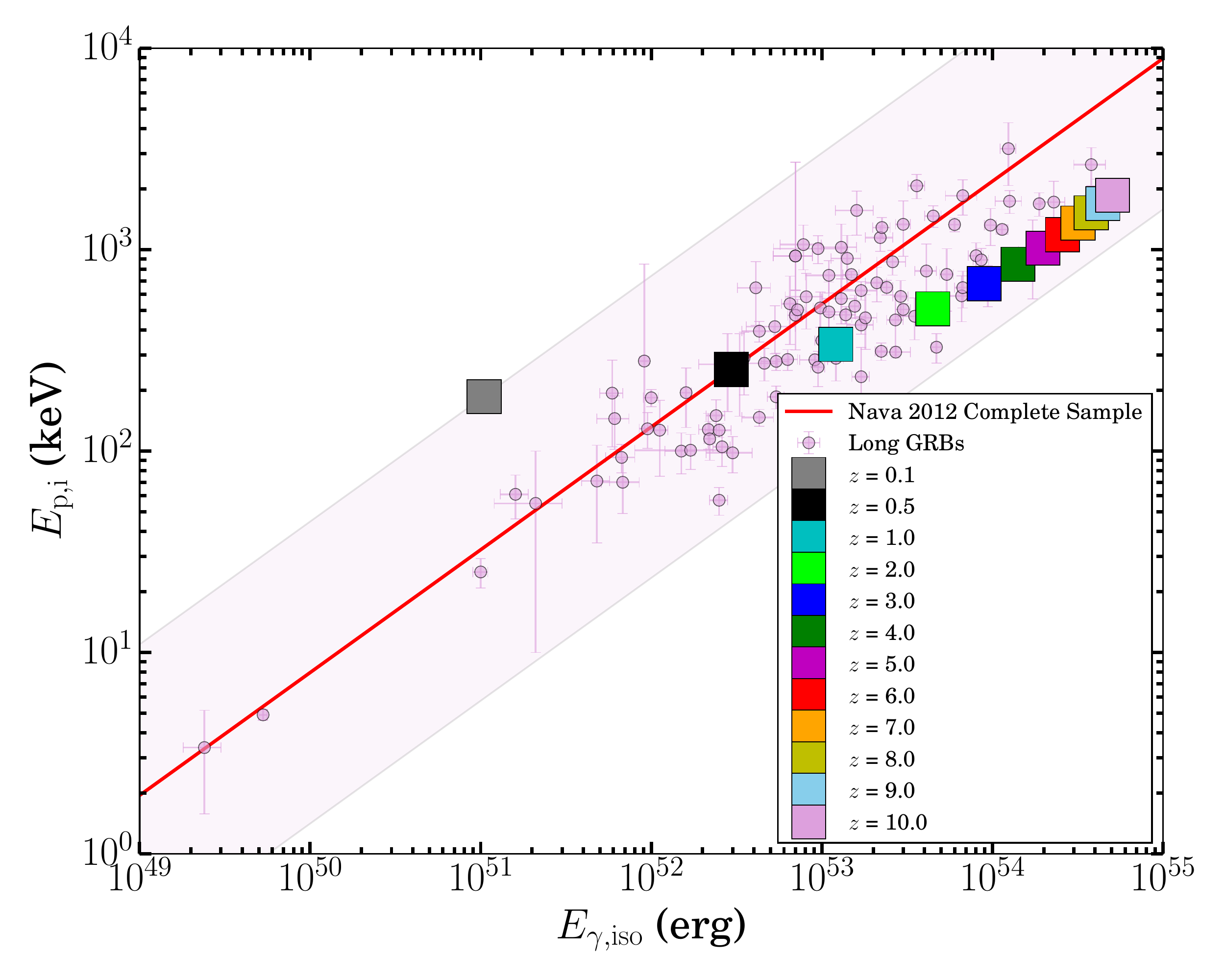}
\caption{Amati correlation for \thisgrb. As there is no redshift measurement is available, we have to vary the redshift from 0.1 to 10. For the comparison, we have also shown the other long bursts studied by \citet{2012MNRAS.421.1256N}. The solid red line shows the best-fit line, and the pink shaded band shows the associated 2-$\sigma$ uncertainty using the study of a sample of long bursts by \citet{2012MNRAS.421.1256N}.}
\label{amati}
\end{figure}

From our deep afterglow/host observations using GTC, no optical counterpart or confirmed associated host galaxy (see section \ref{host SED}) is detected. Therefore, we could not calculate the exact redshift of the burst. To determine the redshift and to examine the high redshift possibility for optical darkness, we utilised prompt emission Amati correlation (the correlation between rest frame \Ep and isotropic gamma-ray energy). We have used time-averaged spectral parameters and bolometric fluence in the source frame for different values of redshift ranging from $z$= 0.1 to $z$= 10. The Amati correlation for \thisgrb at different redshift values is shown in Fig. \ref{amati}. For the comparison, we have also shown the other long bursts studied by \citet{2012MNRAS.421.1256N}. This analysis indicates that the isotropic gamma-ray energy of \thisgrb is such that it would not be excessive even for $z \geq$ 10. The very faint/non-detection of the afterglow in NIR/UV/optical filters also supports a high-$z$ origin. In addition, the prompt emission duration of \thisgrb would be $\sim$ 5 s in the source frame at $z \geq$ 10, consistent with a long GRB. 

However, the high redshift origin of \thisgrb is challenged based on the following argument: considering a high redshift origin of \thisgrb, the soft X-rays photons in the source frame will be shifted out of the XRT energy coverage (0.3-10 \keV). Therefore, even a large column density in the source will result in little attenuation. Furthermore, we utilised the X-ray afterglow spectrum to constrain the redshift of \thisgrb. We fitted the X-ray PC mode spectral data and derived (observer frame) considering $z$ = 0. The measured column density is higher than that of Galactic column density, and this excess column density is useful to estimate the limit on $z$. We have used the following equation to constrain the limit on $z$ \citet{2007AJ....133.2216G}:

\begin{equation}
\centering
\log{(1+z)} < 1.3 - 0.5 \,\log_{10}{(1 + \Delta N_H)},
\end{equation}

where $\Delta N_H$ represents the difference between column density measured considering $z$ = 0 ($N_H \sim 23.06 \times 10^{20} \mathrm{cm^{-2}}$) and Galactic ($N_H \sim 9.05 \times 10^{20} \mathrm{cm^{-2}}$) taken from \citet{2013MNRAS.431..394W}. The above equation indicates that \thisgrb has a redshift value $z<4.15$.

\subsection{\bf Nature of the potential host galaxy of \thisgrb} 
\label{host SED} 
Deeper observations searching the host of \thisgrb found a potential host with an angular separation of $\sim$ 1.1$''$ from the NIR afterglow position (see section \ref{Host Galaxy}). The potential host is a faint ($r\sim 25.26 \pm 0.27\,\rm{mag}$) one with typical $g-i$ colour measured for galaxies at moderate redshift values detected in the optical bands (see Table \ref{optical/NIR observations}). The observed spectrum of the potential host galaxy taken using GTC was very noisy, and there is no emission-like feature up to $z$=1.08, discarding the low-redshift possibility (the non-detection of UV emission also discards the low redshift possibility). So, we executed the SED modelling of the potential galaxy of \thisgrb utilising photometric data and \sw{Prospector} software. \sw{Prospector} is a python-based stellar population code developed for the host galaxy SED modelling using both photometric and spectroscopic observations \citep{2017ApJ...837..170L, 2021ApJS..254...22J}. In order to model the observed data, \sw{Prospector} applies a library of FSPS (Flexible Stellar Population Synthesis) models \citep{2009ApJ...699..486C}. We used the dynamic nested sampling fitting routine \sw{dynesty} on the observed photometry of the potential host galaxy and calculated the posterior distributions of host galaxy parameters. To determine the stellar population properties of the potential host galaxy, we used the \sw{parametric$\_$sfh} model. This model enabled us to calculate several key host properties, including the stellar mass formed ($M_\star$) in solar mass units, host galaxy age ($t$), stellar metallicity ($\log Z/Z_\odot$), dust attenuation (A$_{\rm V}$), and the star formation timescale ($\tau$). In our SED analysis, we used these host galaxy model parameters as free variables. A detailed method of host galaxy SED modelling is described in \citet{2022JApA...43...82G}. Due to the limited number of data points, we also included the limiting mag values of $z$, $H$, and $K_{\rm S}$ filters for the SED analysis. We assumed redshift as a free parameter to constrain the photometric redshift of the potential host galaxy. Fig. \ref{PotentialHG-SED} and Fig. A2 (in the appendix) show the SED fitting and corresponding corner plot of the potential host galaxy of \thisgrb, respectively. We determined the following parameters using SED modelling: stellar mass formed ($M_\star$) = 10.66$^{+0.44}_{-0.54}$, stellar metallicity ($\log Z/Z_\odot$) = -2.27$^{+1.12}_{-0.83}$, age of the galaxy ($t$) = 5.80$^{+4.77}_{-3.66}$, dust attenuation in rest-frame (A$_{\rm V}$) = 0.97$^{+0.34}_{-0.37}$ mag, and star formation timescale ($\tau$) = 1.68$^{+0.82}_{-0.83}$. Considering the redshift 1.88 obtained from SED modelling, we calculated the physical offset $\sim$ 9.5 kpc between the centre of the potential host galaxy and the NIR afterglow position. The measured physical distance is very large compared to those typically observed for long GRBs, suggesting that the galaxy might not be related with \thisgrb. To further explore the nature of the potential galaxy, we calculated the chance coincidence probability of the candidate galaxy following the method described in \citet{2002AJ....123.1111B}. Using the measured brightness of the galaxy in the r-band and the observed offset values, we derived a chance alignment (P$_{cc}$) of about 5 \%. The derived P$_{c}$ value is small but still significantly high, indicating the candidate galaxy to be the host. In literature, authors have used a diverse range of P$_{cc}$ values, that is, $\sim$ 1-10\%, to establish the association of faint galaxies with such transients. For example, \citet{2002AJ....123.1111B, 2010ApJ...722.1946B, 2022MNRAS.515.4890O}, use Pcc$<$10\%, \citet{2013ApJ...769...56F} use Pcc$<$5\%, and \citet{2014MNRAS.437.1495T} use 1-2\%, to identify if a galaxy is related to afterglow or not. In light of the above analysis, it is hard to decipher whether the observed candidate galaxy is the host of \thisgrb.
	    
\begin{figure*}
\centering
\includegraphics[scale=0.3]{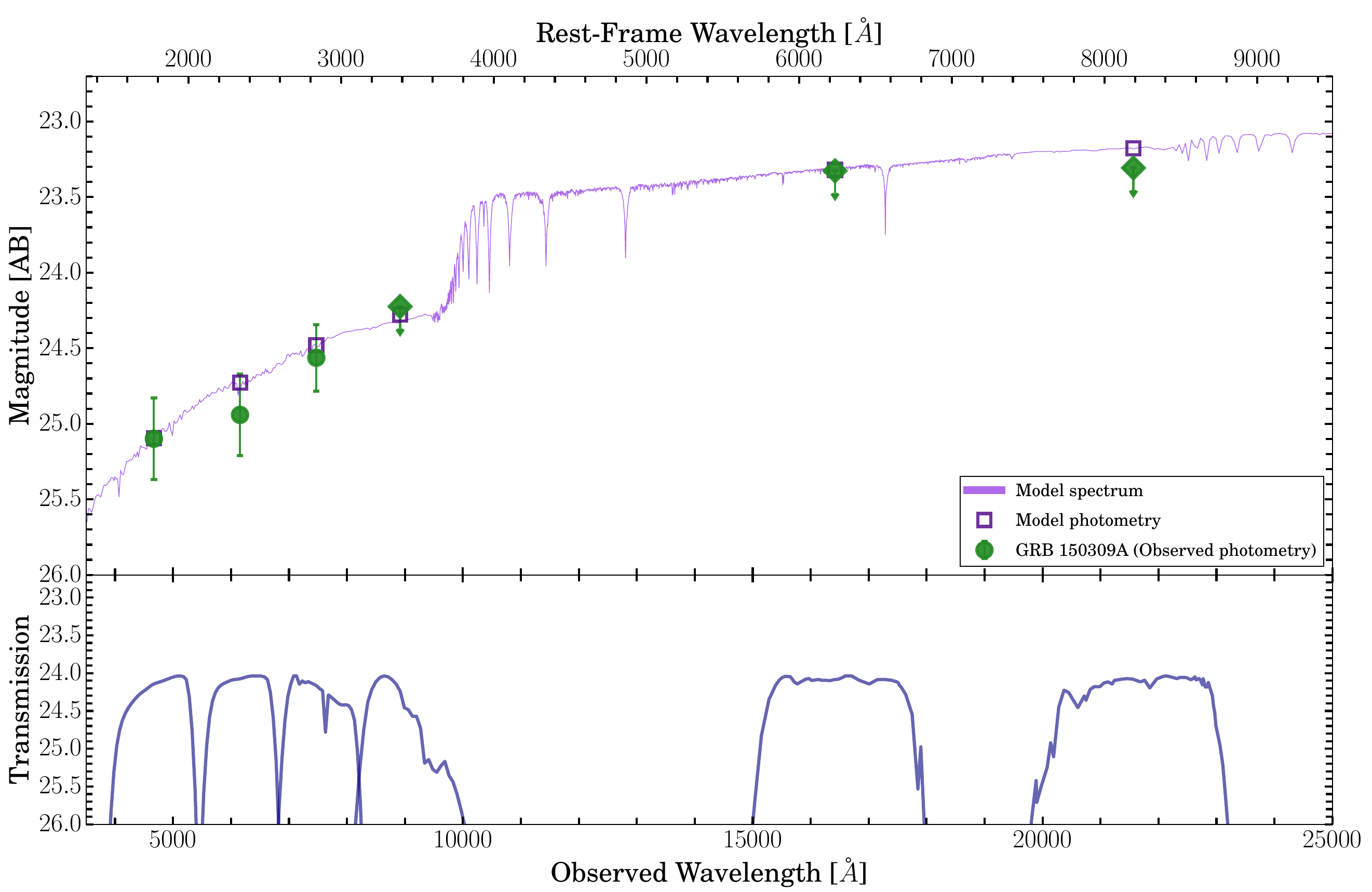}
\caption{Spectral energy distribution modelling of the potential host galaxy (from the $g$-band to the $K_{\rm S}$ band) of \thisgrb using \sw{Prospector} software. The SED fitting shows no evidence of internal reddening. The bottom plot shows the transmission curves of the corresponding filters.}
\label{PotentialHG-SED}
\end{figure*}

\section{Conclusions} 
\label{conclusiones} 

We investigated detailed prompt emission properties of \thisgrb detected by \swift and \fermi. The high energy light curve of \thisgrb consists of a rarely observed two-episodic emission phase intersected by a quiescent gap, the first brighter main episode prior to the second fainter and softer episode. The episode-wise time-averaged as well as time-resolved spectral analysis of high energy observations of \thisgrb exhibit intensity-tracking peak energy \Ep evolution. The evolution of the low-energy index ($\alpha_{\rm pt}$) of \sw{Band} function overshoot the synchrotron limits in early temporal bins of the first episode posing challenges for the thin shell synchrotron origin and demands for photospheric contribution; however, more complex synchrotron emission models (such as synchrotron emission in decaying magnetic field or time-dependent cooling of electrons) may produce harder $\alpha_{\rm pt}$. Our detailed time-resolved spectral analysis of high energy data helps us to understand the radiation physics and jet composition of \thisgrb. The evolution of $\alpha_{\rm pt}$ indicates a hybrid jet composition consisting of a matter-dominated fireball and magnetic-dominated Poynting flux. Considering the value of $E_{\rm \gamma, iso}$ for an assumed value of $z$=2, and derived value of $E_{\rm K, iso}$, we constrained the radiative efficiency ($\eta$= $E_{\rm \gamma, iso}$/($E_{\rm \gamma, iso}$+ $E_{\rm K, iso}$)) $<$ 0.4, typically similar to those found in case of other long GRBs.

Detection of the very red afterglow is crucial to explore the early Universe as they are probes to look for very high redshifts or test beds studying dusty environments surrounding GRBs. In this work, we apprise the discovery of a very red counterpart ($K_{\rm S}$-band) of \thisgrb $\sim$ 5.2 hours post burst with the CIRCE instrument mounted at the GTC, but it was not detected in any bluer filters of \swift UVOT/BOOTES, suggesting for a very high redshift origin. However, our present analysis discarded this possibility based on a few arguments, including spectral analysis of X-ray afterglow, constrain $z <$ 4.15; SED modelling of the potential nearby galaxy constraining moderate redshift values and offset analysis between the centroid of the potential host to the location of the afterglow etc. Furthermore, we also performed the SED modelling of the potential host galaxy. Our results demand a relatively lower value of extinction and redshift along with typical physical parameters with a rather large physical offset from the galaxy's centre. The considerable difference between A$_{\rm V}$ values obtained from the host galaxy SED modelling and the one estimated from afterglow SED analysis of \thisgrb is an indicative of local dust surrounding the progenitor or anomalous extinction within the host galaxy as described in \citet{2009AJ....138.1690P, 2013ApJ...778..128P}. Based on our analysis of the potential host, it is hard to decipher whether the observed candidate galaxy is the host of \thisgrb.

Our analysis of the afterglow SED shows that significant reddening is required to describe the observed $K_{\rm S}$ band (NIR) afterglow of \thisgrb, assuming the cooling frequency of external synchrotron model is beyond the optical/NIR and X-rays frequencies. Following this method, we calculated the $K_{\rm S}$ band host reddening (A$_{K_{\rm S}}$) = 3.60$^{+0.80}_{-0.76}$ mag, equivalent to A$_{\rm V}$ =  34.67$^{+7.70}_{-7.32}$ mag. Our analysis also indicates that \thisgrb is one of the most intense dark GRBs detected so far. Our results suggest that the environment of \thisgrb demands a high extinction towards the line of sight. Hence, dust obscuration is the most probable reason for the optical darkness of \thisgrb. 

The synergy between state-of-the-art NIR cameras such as CIRCE and the largest diameter optical telescope available so far (the GTC), along with the recently launched James Webb Space Telescope (JWST), makes an ideal combination for studying the population of dark GRBs to a great extent and determining if extinction in the host galaxy or very high redshift is the reason for a significant fraction of the afterglows being beyond the reach of optical telescopes in about 20\% of events \citep{2011A&A...526A..30G, 2008MNRAS.388.1743T, 2011A&A...532A..48D}. Furthermore, the community is developing several larger optical telescopes, for example, the Extremely Large Telescope (ELT) and Thirty Meter Telescope (TMT). Our study provides insights for future observations of similar fainter afterglows of dark GRBs using upcoming larger telescopes \citep{2013JApA...34..157P}. 

\begin{acknowledgements}  
We thank the anonymous referee for providing constructive and positive comments. This work is based partly on observations carried out with the 10.4m Gran Telescopio Canarias (GTC) at the Spanish island of La Palma. This work has partially made use of data products from the Two Micron All Sky Survey (2MASS), which is a joint project of the Univ. of Massachusetts and the Infrared Processing and Analysis Center/California Institute of Technology, funded by the National Aeronautics and Space Administration and the National Science Foundation. This work has made use of data obtained from the High Energy Astrophysics Science Archive Research Center (HEASARC) and the Leicester Database and Archive Service (LEDAS), provided by NASA's Goddard Space Flight Center and the Department of Physics and Astronomy, Leicester University, UK, respectively. RG and SBP acknowledge the financial support of ISRO under AstroSat archival Data utilisation program (DS$\_$2B-13013(2)/1/2021-Sec.2). AJCT acknowledges support from the Spanish Ministry project PID2020-118491GB-I00 and Junta de Andalucia grant P20\_010168 and from the Severo Ochoa grant CEX2021-001131-S funded by MCIN/AEI/ 10.13039/501100011033. MCG acknowledges support from the Ram\'on y Cajal Fellowship RYC2019-026465-I (funded by the MCIN/AEI /10.13039/501100011033 and the European Social Funding). YDH acknowledges support under the additional funding from the RYC2019-026465-I. RS-R acknowledges support under the CSIC-MURALES project with reference 20215AT009. MG acknowledges the Academy of Finland project No. 325806. Research in UrFU is supported by the Program of Development within the Priority-2030.
\end{acknowledgements} 
  
\bibliographystyle{aa} 
\bibliography{GRB150309A}

\begin{thebibliography}{89}
\expandafter\ifx\csname natexlab\endcsname\relax\def\natexlab#1{#1}\fi

\bibitem[{Abbott {et~al.}(2017)Abbott, Abbott, Abbott, Acernese, Ackley, Adams,
  Adams, Addesso, Adhikari, Adya, \& et~al.}]{Abbott_2017}
Abbott, B.~P., Abbott, R., Abbott, T.~D., {et~al.} 2017, The Astrophysical
  Journal, 848, L13

\bibitem[{{Ahumada} {et~al.}(2021){Ahumada}, {Singer}, {Anand}, {Coughlin},
  {Kasliwal}, {Ryan}, {Andreoni}, {Cenko}, {Fremling}, {Kumar}, {Pang},
  {Burns}, {Cunningham}, {Dichiara}, {Dietrich}, {Svinkin}, {Almualla},
  {Castro-Tirado}, {De}, {Dunwoody}, {Gatkine}, {Hammerstein}, {Iyyani},
  {Mangan}, {Perley}, {Purkayastha}, {Bellm}, {Bhalerao}, {Bolin}, {Bulla},
  {Cannella}, {Chandra}, {Duev}, {Frederiks}, {Gal-Yam}, {Graham}, {Ho},
  {Hurley}, {Karambelkar}, {Kool}, {Kulkarni}, {Mahabal}, {Masci}, {McBreen},
  {Pandey}, {Reusch}, {Ridnaia}, {Rosnet}, {Rusholme}, {Carracedo}, {Smith},
  {Soumagnac}, {Stein}, {Troja}, {Tsvetkova}, {Walters}, \&
  {Valeev}}]{2021NatAs...5..917A}
{Ahumada}, T., {Singer}, L.~P., {Anand}, S., {et~al.} 2021, Nature Astronomy,
  5, 917

\bibitem[{{Alam} {et~al.}(2015){Alam}, {Albareti}, {Allende Prieto}, {Anders},
  {Anderson}, {Anderton}, {Andrews}, {Armengaud}, {Aubourg}, {Bailey}, \&
  et~al.}]{ala15}
{Alam}, S., {Albareti}, F.~D., {Allende Prieto}, C., {et~al.} 2015, \apjs, 219,
  12

\bibitem[{{Amati}(2006)}]{2006MNRAS.372..233A}
{Amati}, L. 2006, \mnras, 372, 233

\bibitem[{{Arnaud}(1996)}]{1996ASPC..101...17A}
{Arnaud}, K.~A. 1996, in Astronomical Society of the Pacific Conference Series,
  Vol. 101, Astronomical Data Analysis Software and Systems V, ed. G.~H.
  {Jacoby} \& J.~{Barnes}, 17

\bibitem[{{Barthelmy} {et~al.}(2005){Barthelmy}, {Barbier}, {Cummings},
  {Fenimore}, {Gehrels}, {Hullinger}, {Krimm}, {Markwardt}, {Palmer},
  {Parsons}, {Sato}, {Suzuki}, {Takahashi}, {Tashiro}, \& {Tueller}}]{bar05}
{Barthelmy}, S.~D., {Barbier}, L.~M., {Cummings}, J.~R., {et~al.} 2005, Space
  Science Reviews, 120, 143

\bibitem[{{Basak} \& {Rao}(2013)}]{2013MNRAS.436.3082B}
{Basak}, R. \& {Rao}, A.~R. 2013, \mnras, 436, 3082

\bibitem[{{Belczynski} {et~al.}(2002){Belczynski}, {Bulik}, \&
  {Rudak}}]{2002ApJ...571..394B}
{Belczynski}, K., {Bulik}, T., \& {Rudak}, B. 2002, \apj, 571, 394

\bibitem[{{Berger}(2010)}]{2010ApJ...722.1946B}
{Berger}, E. 2010, \apj, 722, 1946

\bibitem[{{Bloom} {et~al.}(1998){Bloom}, {Djorgovski}, {Kulkarni}, \&
  {Frail}}]{1998ApJ...507L..25B}
{Bloom}, J.~S., {Djorgovski}, S.~G., {Kulkarni}, S.~R., \& {Frail}, D.~A. 1998,
  \apjl, 507, L25

\bibitem[{{Bloom} {et~al.}(2002){Bloom}, {Kulkarni}, \&
  {Djorgovski}}]{2002AJ....123.1111B}
{Bloom}, J.~S., {Kulkarni}, S.~R., \& {Djorgovski}, S.~G. 2002, \aj, 123, 1111

\bibitem[{{Burgess}(2014)}]{2014MNRAS.445.2589B}
{Burgess}, J.~M. 2014, \mnras, 445, 2589

\bibitem[{{Burrows} {et~al.}(2005){Burrows}, {Hill}, {Nousek}, {Kennea},
  {Wells}, {Osborne}, {Abbey}, {Beardmore}, {Mukerjee}, {Short}, {Chincarini},
  {Campana}, {Citterio}, {Moretti}, {Pagani}, {Tagliaferri}, {Giommi},
  {Capalbi}, {Tamburelli}, {Angelini}, {Cusumano}, {Br{\"a}uninger}, {Burkert},
  \& {Hartner}}]{2005SSRv..120..165B}
{Burrows}, D.~N., {Hill}, J.~E., {Nousek}, J.~A., {et~al.} 2005, \ssr, 120, 165

\bibitem[{{Caballero-Garc{\'\i}a} {et~al.}(2023){Caballero-Garc{\'\i}a},
  {Gupta}, {Pandey}, {Oates}, {Marisaldi}, {Ramsli}, {Hu}, {Castro-Tirado},
  {S{\'a}nchez-Ram{\'\i}rez}, {Connell}, {Christiansen}, {Ror}, {Aryan}, {Bai},
  {Castro-Tirado}, {Fan}, {Fern{\'a}ndez-Garc{\'\i}a}, {Kumar}, {Lindanger},
  {Mezentsev}, {Navarro-Gonz{\'a}lez}, {Neubert}, {{\O}stgaard},
  {P{\'e}rez-Garc{\'\i}a}, {Reglero}, {Sarria}, {Sun}, {Xiong}, {Yang}, {Yang},
  \& {Zhang}}]{2023MNRAS.519.3201C}
{Caballero-Garc{\'\i}a}, M.~D., {Gupta}, R., {Pandey}, S.~B., {et~al.} 2023,
  \mnras, 519, 3201

\bibitem[{{Cano} {et~al.}(2017){Cano}, {Wang}, {Dai}, \&
  {Wu}}]{2017AdAst2017E...5C}
{Cano}, Z., {Wang}, S.-Q., {Dai}, Z.-G., \& {Wu}, X.-F. 2017, Advances in
  Astronomy, 2017, 8929054

\bibitem[{{Castro-Tirado} {et~al.}(2007){Castro-Tirado}, {Bremer}, {McBreen},
  {Gorosabel}, {Guziy}, {Fakthullin}, {Sokolov}, {Gonz{\'a}lez Delgado},
  {Bihain}, {Pandey}, {Jel{\'\i}nek}, {de Ugarte Postigo}, {Misra}, {Sagar},
  {Bama}, {Kamble}, {Anupama}, {Licandro}, {P{\'e}rez-Ram{\'\i}rez},
  {Bhattacharya}, {Aceituno}, \& {Neri}}]{2007A&A...475..101C}
{Castro-Tirado}, A.~J., {Bremer}, M., {McBreen}, S., {et~al.} 2007, \aap, 475,
  101

\bibitem[{{Chand} {et~al.}(2020){Chand}, {Banerjee}, {Gupta}, {Dimple}, {Pal},
  {Joshi}, {Zhang}, {Basak}, {Tam}, {Sharma}, {Pand ey}, {Kumar}, \&
  {Yang}}]{2020ApJ...898...42C}
{Chand}, V., {Banerjee}, A., {Gupta}, R., {et~al.} 2020, \apj, 898, 42

\bibitem[{{Conroy} {et~al.}(2009){Conroy}, {Gunn}, \&
  {White}}]{2009ApJ...699..486C}
{Conroy}, C., {Gunn}, J.~E., \& {White}, M. 2009, \apj, 699, 486

\bibitem[{{Cummings} {et~al.}(2015){Cummings}, {Barthelmy}, {Page}, {Palmer},
  \& {Ukwatta}}]{Cummings15}
{Cummings}, J.~R., {Barthelmy}, S.~D., {Page}, K.~L., {Palmer}, D.~M., \&
  {Ukwatta}, T.~N. 2015, GRB Coordinates Network, 17553, 1

\bibitem[{{De Pasquale} {et~al.}(2003){De Pasquale}, {Piro}, {Perna}, {Costa},
  {Feroci}, {Gandolfi}, {in 't Zand}, {Nicastro}, {Frontera}, {Antonelli},
  {Fiore}, \& {Stratta}}]{DePasquale03}
{De Pasquale}, M., {Piro}, L., {Perna}, R., {et~al.} 2003, \apj, 592, 1018

\bibitem[{{D'Elia} \& {Stratta}(2011)}]{2011A&A...532A..48D}
{D'Elia}, V. \& {Stratta}, G. 2011, \aap, 532, A48

\bibitem[{{Della Valle} {et~al.}(2006){Della Valle}, {Chincarini}, {Panagia},
  {Tagliaferri}, {Malesani}, {Testa}, {Fugazza}, {Campana}, {Covino},
  {Mangano}, {Antonelli}, {D'Avanzo}, {Hurley}, {Mirabel}, {Pellizza},
  {Piranomonte}, \& {Stella}}]{2006Natur.444.1050D}
{Della Valle}, M., {Chincarini}, G., {Panagia}, N., {et~al.} 2006, \nat, 444,
  1050

\bibitem[{{Evans} {et~al.}(2009){Evans}, {Beardmore}, {Page}, {Osborne},
  {O'Brien}, {Willingale}, {Starling}, {Burrows}, {Godet}, {Vetere}, {Racusin},
  {Goad}, {Wiersema}, {Angelini}, {Capalbi}, {Chincarini}, {Gehrels}, {Kennea},
  {Margutti}, {Morris}, {Mountford}, {Pagani}, {Perri}, {Romano}, \&
  {Tanvir}}]{eva09}
{Evans}, P.~A., {Beardmore}, A.~P., {Page}, K.~L., {et~al.} 2009, \mnras, 397,
  1177

\bibitem[{{Evans} {et~al.}(2007){Evans}, {Beardmore}, {Page}, {Tyler},
  {Osborne}, {Goad}, {O'Brien}, {Vetere}, {Racusin}, {Morris}, {Burrows},
  {Capalbi}, {Perri}, {Gehrels}, \& {Romano}}]{eva07}
{Evans}, P.~A., {Beardmore}, A.~P., {Page}, K.~L., {et~al.} 2007, \aap, 469,
  379

\bibitem[{{Fong} {et~al.}(2013){Fong}, {Berger}, {Chornock}, {Margutti},
  {Levan}, {Tanvir}, {Tunnicliffe}, {Czekala}, {Fox}, {Perley}, {Cenko},
  {Zauderer}, {Laskar}, {Persson}, {Monson}, {Kelson}, {Birk}, {Murphy},
  {Servillat}, \& {Anglada}}]{2013ApJ...769...56F}
{Fong}, W., {Berger}, E., {Chornock}, R., {et~al.} 2013, \apj, 769, 56

\bibitem[{{Fynbo} {et~al.}(2001){Fynbo}, {Gorosabel}, {Dall}, {Hjorth},
  {Pedersen}, {Andersen}, {M{\o}ller}, {Holland }, {Smail}, {Kobayashi}, {Rol},
  {Vreeswijk}, {Burud}, {Jensen}, {Thomsen}, {Henden}, {Vrba}, {Canzian},
  {Castro Cer{\'o}n}, {Castro-Tirado}, {Cline}, {Goto}, {Greiner}, {Hanski},
  {Hurley}, {Lund}, {Pursimo}, {{\O}stensen}, {Solheim}, {Tanvir}, \&
  {Terada}}]{Fynbo01}
{Fynbo}, J.~U., {Gorosabel}, J., {Dall}, T.~H., {et~al.} 2001, \aap, 373, 796

\bibitem[{{Genet} \& {Granot}(2009)}]{2009MNRAS.399.1328G}
{Genet}, F. \& {Granot}, J. 2009, \mnras, 399, 1328

\bibitem[{{Goldstein} {et~al.}(2017){Goldstein}, {Veres}, {Burns}, {Briggs},
  {Hamburg}, {Kocevski}, {Wilson-Hodge}, {Preece}, {Poolakkil}, {Roberts},
  {Hui}, {Connaughton}, {Racusin}, {von Kienlin}, {Dal Canton}, {Christensen},
  {Littenberg}, {Siellez}, {Blackburn}, {Broida}, {Bissaldi}, {Cleveland},
  {Gibby}, {Giles}, {Kippen}, {McBreen}, {McEnery}, {Meegan}, {Paciesas}, \&
  {Stanbro}}]{2017ApJ...848L..14G}
{Goldstein}, A., {Veres}, P., {Burns}, E., {et~al.} 2017, \apjl, 848, L14

\bibitem[{{Golenetskii} {et~al.}(2015){Golenetskii}, {Aptekar}, {Frederiks},
  {Pal'Shin}, {Oleynik}, {Ulanov}, {Svinkin}, {Tsvetkova}, {Lysenko}, \&
  {Cline}}]{Golenetskii15}
{Golenetskii}, S., {Aptekar}, R., {Frederiks}, D., {et~al.} 2015, GRB
  Coordinates Network, 17566, 1

\bibitem[{{Golkhou} {et~al.}(2015){Golkhou}, {Butler}, \&
  {Littlejohns}}]{2015ApJ...811...93G}
{Golkhou}, V.~Z., {Butler}, N.~R., \& {Littlejohns}, O.~M. 2015, \apj, 811, 93

\bibitem[{{Greiner} {et~al.}(2011){Greiner}, {Kr{\"u}hler}, {Klose}, {Afonso},
  {Clemens}, {Filgas}, {Hartmann}, {K{\"u}pc{\"u} Yolda{\textcommabelow s}},
  {Nardini}, {Olivares E.}, {Rau}, {Rossi}, {Schady}, \&
  {Updike}}]{2011A&A...526A..30G}
{Greiner}, J., {Kr{\"u}hler}, T., {Klose}, S., {et~al.} 2011, \aap, 526, A30

\bibitem[{{Groot} {et~al.}(1998){Groot}, {Galama}, {van Paradijs},
  {Kouveliotou}, {Wijers}, {Bloom}, {Tanvir}, {Vanderspek}, {Greiner},
  {Castro-Tirado}, {Gorosabel}, {von Hippel}, {Lehnert}, {Kuijken}, {Hoekstra},
  {Metcalfe}, {Howk}, {Conselice}, {Telting}, {Rutten}, {Rhoads}, {Cole},
  {Pisano}, {Naber}, \& {Schwarz}}]{1998ApJ...493L..27G}
{Groot}, P.~J., {Galama}, T.~J., {van Paradijs}, J., {et~al.} 1998, \apjl, 493,
  L27

\bibitem[{{Grupe} {et~al.}(2007){Grupe}, {Nousek}, {vanden Berk}, {Roming},
  {Burrows}, {Godet}, {Osborne}, \& {Gehrels}}]{2007AJ....133.2216G}
{Grupe}, D., {Nousek}, J.~A., {vanden Berk}, D.~E., {et~al.} 2007, \aj, 133,
  2216

\bibitem[{{Gupta} {et~al.}(2021){Gupta}, {Oates}, {Pandey}, {Castro-Tirado},
  {Joshi}, {Hu}, {Valeev}, {Zhang}, {Zhang}, {Kumar}, {Aryan}, {Lien}, {Kumar},
  {Cui}, {Wang}, {Dimple}, {Bhattacharya}, {Sonbas}, {Bai}, {Tello},
  {Gorosabel}, {Castro Cer{\'o}n}, {Porto}, {Misra}, {De Pasquale},
  {Caballero-Garc{\'\i}a}, {Jel{\'\i}nek}, {Kub{\'a}nek}, {Minaev}, {Cunniffe},
  {S{\'a}nchez-Ram{\'\i}rez}, {Guziy}, {Jeong}, {Tiwari}, {Razzaque},
  {Bhalerao}, {Pintado}, {Sokolov}, {Zhao}, {Fan}, \&
  {Xin}}]{2021MNRAS.505.4086G}
{Gupta}, R., {Oates}, S.~R., {Pandey}, S.~B., {et~al.} 2021, \mnras, 505, 4086

\bibitem[{{Gupta} {et~al.}(2022){Gupta}, {Pandey}, {Kumar}, {Aryan}, {Ror},
  {Sharma}, {Misra}, {Castro-Tirado}, \& {Tiwari}}]{2022JApA...43...82G}
{Gupta}, R., {Pandey}, S.~B., {Kumar}, A., {et~al.} 2022, Journal of
  Astrophysics and Astronomy, 43, 82

\bibitem[{{Hashimoto} {et~al.}(2010){Hashimoto}, {Ohta}, {Aoki}, {Tanaka},
  {Yabe}, {Kawai}, {Aoki}, {Furusawa}, {Hattori}, {Iye}, {Kawabata},
  {Kobayashi}, {Komiyama}, {Kosugi}, {Minowa}, {Mizumoto}, {Niino}, {Nomoto},
  {Noumaru}, {Ogasawara}, {Pyo}, {Sakamoto}, {Sekiguchi}, {Shirasaki},
  {Suzuki}, {Tajitsu}, {Takata}, {Tamagawa}, {Terada}, {Totani}, {Watanabe},
  {Yamada}, \& {Yoshida}}]{2010ApJ...719..378H}
{Hashimoto}, T., {Ohta}, K., {Aoki}, K., {et~al.} 2010, \apj, 719, 378

\bibitem[{{Heger} {et~al.}(2003){Heger}, {Fryer}, {Woosley}, {Langer}, \&
  {Hartmann}}]{2003ApJ...591..288H}
{Heger}, A., {Fryer}, C.~L., {Woosley}, S.~E., {Langer}, N., \& {Hartmann},
  D.~H. 2003, \apj, 591, 288

\bibitem[{{Jakobsson} {et~al.}(2004){Jakobsson}, {Hjorth}, {Fynbo}, {Watson},
  {Pedersen}, {Bj{\"o}rnsson}, \& {Gorosabel}}]{Jakobsson04}
{Jakobsson}, P., {Hjorth}, J., {Fynbo}, J.~P.~U., {et~al.} 2004, \apjl, 617,
  L21

\bibitem[{{Jaunsen} {et~al.}(2008){Jaunsen}, {Rol}, {Watson}, {Malesani},
  {Fynbo}, {Milvang-Jensen}, {Hjorth}, {Vreeswijk}, {Ovaldsen}, {Wiersema},
  {Tanvir}, {Gorosabel}, {Levan}, {Schirmer}, \&
  {Castro-Tirado}}]{2008ApJ...681..453J}
{Jaunsen}, A.~O., {Rol}, E., {Watson}, D.~J., {et~al.} 2008, \apj, 681, 453

\bibitem[{{Johnson} {et~al.}(2021){Johnson}, {Leja}, {Conroy}, \&
  {Speagle}}]{2021ApJS..254...22J}
{Johnson}, B.~D., {Leja}, J., {Conroy}, C., \& {Speagle}, J.~S. 2021, \apjs,
  254, 22

\bibitem[{{Kouveliotou} {et~al.}(1993){Kouveliotou}, {Meegan}, {Fishman},
  {Bhat}, {Briggs}, {Koshut}, {Paciesas}, \& {Pendleton}}]{kou93}
{Kouveliotou}, C., {Meegan}, C.~A., {Fishman}, G.~J., {et~al.} 1993, \apjl,
  413, L101

\bibitem[{{Kumar} \& {Panaitescu}(2000)}]{2000ApJ...541L..51K}
{Kumar}, P. \& {Panaitescu}, A. 2000, \apjl, 541, L51

\bibitem[{{Kumar} \& {Zhang}(2015)}]{2015PhR...561....1K}
{Kumar}, P. \& {Zhang}, B. 2015, \physrep, 561, 1

\bibitem[{{Lamb} \& {Reichart}(2000)}]{Lamb00}
{Lamb}, D.~Q. \& {Reichart}, D.~E. 2000, \apj, 536, 1

\bibitem[{{Lan} {et~al.}(2018){Lan}, {L{\"u}}, {Zhong}, {Zhang}, {Rice},
  {Cheng}, {Du}, {Li}, {Lin}, {Lu}, \& {Liang}}]{2018ApJ...862..155L}
{Lan}, L., {L{\"u}}, H.-J., {Zhong}, S.-Q., {et~al.} 2018, \apj, 862, 155

\bibitem[{{Leja} {et~al.}(2017){Leja}, {Johnson}, {Conroy}, {van Dokkum}, \&
  {Byler}}]{2017ApJ...837..170L}
{Leja}, J., {Johnson}, B.~D., {Conroy}, C., {van Dokkum}, P.~G., \& {Byler}, N.
  2017, \apj, 837, 170

\bibitem[{{Li} {et~al.}(2020){Li}, {Ryde}, {Pe'er}, {Yu}, \&
  {Acuner}}]{2020arXiv201203038L}
{Li}, L., {Ryde}, F., {Pe'er}, A., {Yu}, H.-F., \& {Acuner}, Z. 2020, arXiv
  e-prints, arXiv:2012.03038

\bibitem[{{Li} {et~al.}(2018){Li}, {Wu}, {Lei}, {Dai}, {Liang}, \&
  {Ryde}}]{2018ApJS..236...26L}
{Li}, L., {Wu}, X.-F., {Lei}, W.-H., {et~al.} 2018, \apjs, 236, 26

\bibitem[{{Littlejohns} {et~al.}(2015){Littlejohns}, {Butler}, {Cucchiara},
  {Watson}, {Fox}, {Lee}, {Kutyrev}, {Richer}, {Klein}, {Prochaska}, {Bloom},
  {Troja}, {Ramirez-Ruiz}, {de Diego}, {Georgiev}, {Gonz{\'a}lez},
  {Rom{\'a}n-Z{\'u}{\~n}iga}, {Gehrels}, \& {Moseley}}]{2015MNRAS.449.2919L}
{Littlejohns}, O.~M., {Butler}, N.~R., {Cucchiara}, A., {et~al.} 2015, \mnras,
  449, 2919

\bibitem[{{MacFadyen} \& {Woosley}(1999)}]{1999ApJ...524..262M}
{MacFadyen}, A.~I. \& {Woosley}, S.~E. 1999, \apj, 524, 262

\bibitem[{{Martin-Carrillo} {et~al.}(2014){Martin-Carrillo}, {Hanlon},
  {Topinka}, {LaCluyz{\'e}}, {Savchenko}, {Kann}, {Trotter}, {Covino},
  {Kr{\"u}hler}, {Greiner}, {McGlynn}, {Murphy}, {Tisdall}, {Meehan}, {Wade},
  {McBreen}, {Reichart}, {Fugazza}, {Haislip}, {Rossi}, {Schady}, {Elliott}, \&
  {Klose}}]{2014A&A...567A..84M}
{Martin-Carrillo}, A., {Hanlon}, L., {Topinka}, M., {et~al.} 2014, \aap, 567,
  A84

\bibitem[{{Mazaeva} {et~al.}(2015){Mazaeva}, {Inasaridze}, {Makandarashvili},
  {Volnova}, {Molotov}, \& {Pozanenko}}]{2015GCN.17570....1M}
{Mazaeva}, E., {Inasaridze}, R., {Makandarashvili}, S., {et~al.} 2015, GRB
  Coordinates Network, 17570, 1

\bibitem[{{Meegan} {et~al.}(2009){Meegan}, {Lichti}, {Bhat}, {Bissaldi},
  {Briggs}, {Connaughton}, {Diehl}, {Fishman}, {Greiner}, {Hoover}, {van der
  Horst}, {von Kienlin}, {Kippen}, {Kouveliotou}, {McBreen}, {Paciesas},
  {Preece}, {Steinle}, {Wallace}, {Wilson}, \&
  {Wilson-Hodge}}]{2009ApJ...702..791M}
{Meegan}, C., {Lichti}, G., {Bhat}, P.~N., {et~al.} 2009, \apj, 702, 791

\bibitem[{{Melandri} {et~al.}(2012){Melandri}, {Sbarufatti}, {D'Avanzo},
  {Salvaterra}, {Campana}, {Covino}, {Vergani}, {Nava}, {Ghisellini}, {Ghirland
  a}, {Fugazza}, {Mangano}, {Capalbi}, \& {Tagliaferri}}]{2012MNRAS.421.1265M}
{Melandri}, A., {Sbarufatti}, B., {D'Avanzo}, P., {et~al.} 2012, \mnras, 421,
  1265

\bibitem[{{Minaev} \& {Pozanenko}(2020)}]{2020MNRAS.492.1919M}
{Minaev}, P.~Y. \& {Pozanenko}, A.~S. 2020, \mnras, 492, 1919

\bibitem[{{Nava} {et~al.}(2012){Nava}, {Salvaterra}, {Ghirlanda}, {Ghisellini},
  {Campana}, {Covino}, {Cusumano}, {D'Avanzo}, {D'Elia}, {Fugazza}, {Melandri},
  {Sbarufatti}, {Vergani}, \& {Tagliaferri}}]{2012MNRAS.421.1256N}
{Nava}, L., {Salvaterra}, R., {Ghirlanda}, G., {et~al.} 2012, \mnras, 421, 1256

\bibitem[{{Oates} \& {Cummings}(2015)}]{2015GCN.17559....1O}
{Oates}, S.~R. \& {Cummings}, J.~R. 2015, GRB Coordinates Network, 17559, 1

\bibitem[{{Obergaulinger} \& {Aloy}(2020)}]{2020MNRAS.492.4613O}
{Obergaulinger}, M. \& {Aloy}, M.~{\'A}. 2020, \mnras, 492, 4613

\bibitem[{{O'Connor} {et~al.}(2022){O'Connor}, {Troja}, {Dichiara},
  {Beniamini}, {Cenko}, {Kouveliotou}, {Gonz{\'a}lez}, {Durbak}, {Gatkine},
  {Kutyrev}, {Sakamoto}, {S{\'a}nchez-Ram{\'\i}rez}, \&
  {Veilleux}}]{2022MNRAS.515.4890O}
{O'Connor}, B., {Troja}, E., {Dichiara}, S., {et~al.} 2022, \mnras, 515, 4890

\bibitem[{{Panaitescu} \& {Kumar}(2002)}]{2002ApJ...571..779P}
{Panaitescu}, A. \& {Kumar}, P. 2002, \apj, 571, 779

\bibitem[{{Pandey}(2013)}]{2013JApA...34..157P}
{Pandey}, S.~B. 2013, Journal of Astrophysics and Astronomy, 34, 157

\bibitem[{{Pandey} {et~al.}(2003){Pandey}, {Anupama}, {Sagar}, {Bhattacharya},
  {Castro-Tirado}, {Sahu}, {Parihar}, \& {Prabhu}}]{2003A&A...408L..21P}
{Pandey}, S.~B., {Anupama}, G.~C., {Sagar}, R., {et~al.} 2003, \aap, 408, L21

\bibitem[{{Perley} {et~al.}(2009){Perley}, {Cenko}, {Bloom}, {Chen}, {Butler},
  {Kocevski}, {Prochaska}, {Brodwin}, {Glazebrook}, {Kasliwal}, {Kulkarni},
  {Lopez}, {Ofek}, {Pettini}, {Soderberg}, \& {Starr}}]{2009AJ....138.1690P}
{Perley}, D.~A., {Cenko}, S.~B., {Bloom}, J.~S., {et~al.} 2009, \aj, 138, 1690

\bibitem[{{Perley} {et~al.}(2016){Perley}, {Kr{\"u}hler}, {Schulze}, {de Ugarte
  Postigo}, {Hjorth}, {Berger}, {Cenko}, {Chary}, {Cucchiara}, {Ellis}, {Fong},
  {Fynbo}, {Gorosabel}, {Greiner}, {Jakobsson}, {Kim}, {Laskar}, {Levan},
  {Micha{\l}owski}, {Milvang-Jensen}, {Tanvir}, {Th{\"o}ne}, \&
  {Wiersema}}]{Perleya16}
{Perley}, D.~A., {Kr{\"u}hler}, T., {Schulze}, S., {et~al.} 2016, \apj, 817, 7

\bibitem[{{Perley} {et~al.}(2013){Perley}, {Levan}, {Tanvir}, {Cenko}, {Bloom},
  {Hjorth}, {Kr{\"u}hler}, {Filippenko}, {Fruchter}, {Fynbo}, {Jakobsson},
  {Kalirai}, {Milvang-Jensen}, {Morgan}, {Prochaska}, \&
  {Silverman}}]{2013ApJ...778..128P}
{Perley}, D.~A., {Levan}, A.~J., {Tanvir}, N.~R., {et~al.} 2013, \apj, 778, 128

\bibitem[{{Perna} \& {Belczynski}(2002)}]{2002ApJ...570..252P}
{Perna}, R. \& {Belczynski}, K. 2002, \apj, 570, 252

\bibitem[{{Piran}(2004)}]{2004RvMP...76.1143P}
{Piran}, T. 2004, Reviews of Modern Physics, 76, 1143

\bibitem[{{Rastinejad} {et~al.}(2022){Rastinejad}, {Gompertz}, {Levan}, {Fong},
  {Nicholl}, {Lamb}, {Malesani}, {Nugent}, {Oates}, {Tanvir}, {de Ugarte
  Postigo}, {Kilpatrick}, {Moore}, {Metzger}, {Ravasio}, {Rossi}, {Schroeder},
  {Jencson}, {Sand}, {Smith}, {Ag{\"u}{\'\i} Fern{\'a}ndez}, {Berger},
  {Blanchard}, {Chornock}, {Cobb}, {De Pasquale}, {Fynbo}, {Izzo}, {Kann},
  {Laskar}, {Marini}, {Paterson}, {Escorial}, {Sears}, \&
  {Th{\"o}ne}}]{2022Natur.612..223R}
{Rastinejad}, J.~C., {Gompertz}, B.~P., {Levan}, A.~J., {et~al.} 2022, \nat,
  612, 223

\bibitem[{{Roberts} \& {Stanbro}(2015)}]{Roberts15}
{Roberts}, O.~J. \& {Stanbro}, M. 2015, GRB Coordinates Network, 17561, 1

\bibitem[{{Roming} {et~al.}(2005){Roming}, {Kennedy}, {Mason}, {Nousek}, {Ahr},
  {Bingham}, {Broos}, {Carter}, {Hancock}, {Huckle}, {Hunsberger}, {Kawakami},
  {Killough}, {Koch}, {McLelland}, {Smith}, {Smith}, {Soto}, {Boyd},
  {Breeveld}, {Holland}, {Ivanushkina}, {Pryzby}, {Still}, \&
  {Stock}}]{2005SSRv..120...95R}
{Roming}, P. W.~A., {Kennedy}, T.~E., {Mason}, K.~O., {et~al.} 2005, \ssr, 120,
  95

\bibitem[{{Rossi} {et~al.}(2022){Rossi}, {Rothberg}, {Palazzi}, {Kann},
  {D'Avanzo}, {Amati}, {Klose}, {Perego}, {Pian}, {Guidorzi}, {Pozanenko},
  {Savaglio}, {Stratta}, {Agapito}, {Covino}, {Cusano}, {D'Elia}, {De
  Pasquale}, {Della Valle}, {Kuhn}, {Izzo}, {Loffredo}, {Masetti}, {Melandri},
  {Minaev}, {Guelbenzu}, {Paris}, {Paiano}, {Plantet}, {Rossi}, {Salvaterra},
  {Schulze}, {Veillet}, \& {Volnova}}]{2022ApJ...932....1R}
{Rossi}, A., {Rothberg}, B., {Palazzi}, E., {et~al.} 2022, \apj, 932, 1

\bibitem[{{Rossi} {et~al.}(2011){Rossi}, {Schulze}, {Klose}, {Kann}, {Rau},
  {Krimm}, {J{\'o}hannesson}, {Panaitescu}, {Yuan}, {Ferrero}, {Kr{\"u}hler},
  {Greiner}, {Schady}, {Pand ey}, {Amati}, {Afonso}, {Akerlof}, {Arnold},
  {Clemens}, {Filgas}, {Hartmann}, {K{\"u}pc{\"u} Yolda{\textcommabelow s}},
  {McBreen}, {McKay}, {Nicuesa Guelbenzu}, {Olivares}, {Paciesas}, {Rykoff},
  {Szokoly}, {Updike}, \& {Yolda{\textcommabelow s}}}]{2011A&A...529A.142R}
{Rossi}, A., {Schulze}, S., {Klose}, S., {et~al.} 2011, \aap, 529, A142

\bibitem[{{Rumyantsev} {et~al.}(2015){Rumyantsev}, {Antoniuk}, {Mazaeva}, \&
  {Pozanenko}}]{2015GCN.17565....1R}
{Rumyantsev}, V., {Antoniuk}, K., {Mazaeva}, E., \& {Pozanenko}, A. 2015, GRB
  Coordinates Network, 17565, 1

\bibitem[{{Sari} {et~al.}(1998){Sari}, {Piran}, \& {Narayan}}]{Sari98}
{Sari}, R., {Piran}, T., \& {Narayan}, R. 1998, \apjl, 497, L17

\bibitem[{{Scargle} {et~al.}(2013){Scargle}, {Norris}, {Jackson}, \&
  {Chiang}}]{2013arXiv1304.2818S}
{Scargle}, J.~D., {Norris}, J.~P., {Jackson}, B., \& {Chiang}, J. 2013, arXiv
  e-prints, arXiv:1304.2818

\bibitem[{{Schlafly} \& {Finkbeiner}(2011)}]{sch11}
{Schlafly}, E.~F. \& {Finkbeiner}, D.~P. 2011, \apj, 737, 103

\bibitem[{{Stanek} {et~al.}(2003){Stanek}, {Matheson}, {Garnavich}, {Martini},
  {Berlind}, {Caldwell}, {Challis}, {Brown}, {Schild}, {Krisciunas}, {Calkins},
  {Lee}, {Hathi}, {Jansen}, {Windhorst}, {Echevarria}, {Eisenstein}, {Pindor},
  {Olszewski}, {Harding}, {Holland}, \& {Bersier}}]{2003ApJ...591L..17S}
{Stanek}, K.~Z., {Matheson}, T., {Garnavich}, P.~M., {et~al.} 2003, \apjl, 591,
  L17

\bibitem[{{Tanvir} {et~al.}(2008){Tanvir}, {Levan}, {Rol}, {Starling},
  {Gorosabel}, {Priddey}, {Malesani}, {Jakobsson}, {O'Brien}, {Jaunsen},
  {Hjorth}, {Fynbo}, {Melandri}, {Gomboc}, {Milvang-Jensen}, {Fruchter},
  {Jarvis}, {Fernandes}, \& {Wold}}]{2008MNRAS.388.1743T}
{Tanvir}, N.~R., {Levan}, A.~J., {Rol}, E., {et~al.} 2008, \mnras, 388, 1743

\bibitem[{{Taylor} {et~al.}(1998){Taylor}, {Frail}, {Kulkarni}, {Shepherd},
  {Feroci}, \& {Frontera}}]{Taylor98}
{Taylor}, G.~B., {Frail}, D.~A., {Kulkarni}, S.~R., {et~al.} 1998, \apjl, 502,
  L115

\bibitem[{{Troja} {et~al.}(2022){Troja}, {Fryer}, {O'Connor}, {Ryan},
  {Dichiara}, {Kumar}, {Ito}, {Gupta}, {Wollaeger}, {Norris}, {Kawai},
  {Butler}, {Aryan}, {Misra}, {Hosokawa}, {Murata}, {Niwano}, {Pandey},
  {Kutyrev}, {van Eerten}, {Chase}, {Hu}, {Caballero-Garcia}, \&
  {Castro-Tirado}}]{2022Natur.612..228T}
{Troja}, E., {Fryer}, C.~L., {O'Connor}, B., {et~al.} 2022, \nat, 612, 228

\bibitem[{{Tunnicliffe} {et~al.}(2014){Tunnicliffe}, {Levan}, {Tanvir},
  {Rowlinson}, {Perley}, {Bloom}, {Cenko}, {O'Brien}, {Cobb}, {Wiersema},
  {Malesani}, {de Ugarte Postigo}, {Hjorth}, {Fynbo}, \&
  {Jakobsson}}]{2014MNRAS.437.1495T}
{Tunnicliffe}, R.~L., {Levan}, A.~J., {Tanvir}, N.~R., {et~al.} 2014, \mnras,
  437, 1495

\bibitem[{Ukwatta {et~al.}(2010)Ukwatta, Stamatikos, Dhuga, Sakamoto,
  Barthelmy, Eskandarian, Gehrels, Maximon, Norris, \& Parke}]{Ukwatta_2010}
Ukwatta, T.~N., Stamatikos, M., Dhuga, K.~S., {et~al.} 2010, The Astrophysical
  Journal, 711, 1073–1086

\bibitem[{{van der Horst} {et~al.}(2009){van der Horst}, {Kouveliotou},
  {Gehrels}, {Rol}, {Wijers}, {Cannizzo}, {Racusin}, \&
  {Burrows}}]{2009ApJ...699.1087V}
{van der Horst}, A.~J., {Kouveliotou}, C., {Gehrels}, N., {et~al.} 2009, \apj,
  699, 1087

\bibitem[{{Vianello} {et~al.}(2018){Vianello}, {Gill}, {Granot}, {Omodei},
  {Cohen-Tanugi}, \& {Longo}}]{2018ApJ...864..163V}
{Vianello}, G., {Gill}, R., {Granot}, J., {et~al.} 2018, \apj, 864, 163

\bibitem[{{Vianello} {et~al.}(2015){Vianello}, {Lauer}, {Younk}, {Tibaldo},
  {Burgess}, {Ayala}, {Harding}, {Hui}, {Omodei}, \&
  {Zhou}}]{2015arXiv150708343V}
{Vianello}, G., {Lauer}, R.~J., {Younk}, P., {et~al.} 2015, arXiv e-prints,
  arXiv:1507.08343

\bibitem[{{Willingale} {et~al.}(2013){Willingale}, {Starling}, {Beardmore},
  {Tanvir}, \& {O'Brien}}]{2013MNRAS.431..394W}
{Willingale}, R., {Starling}, R.~L.~C., {Beardmore}, A.~P., {Tanvir}, N.~R., \&
  {O'Brien}, P.~T. 2013, \mnras, 431, 394

\bibitem[{{Woosley} \& {Bloom}(2006)}]{2006ARA&A..44..507W}
{Woosley}, S.~E. \& {Bloom}, J.~S. 2006, \araa, 44, 507

\bibitem[{{Yang} {et~al.}(2022){Yang}, {Ai}, {Zhang}, {Zhang}, {Liu}, {Wang},
  {Yang}, {Yin}, {Li}, \& {L{\"u}}}]{2022Natur.612..232Y}
{Yang}, J., {Ai}, S., {Zhang}, B.-B., {et~al.} 2022, \nat, 612, 232

\bibitem[{{Zhang} {et~al.}(2021){Zhang}, {Liu}, {Peng}, {Li}, {L{\"u}}, {Yang},
  {Yang}, {Yang}, {Meng}, {Zou}, {Ye}, {Wang}, {Mao}, {Zhao}, {Bai},
  {Castro-Tirado}, {Hu}, {Dai}, {Liang}, \& {Zhang}}]{2021NatAs...5..911Z}
{Zhang}, B.~B., {Liu}, Z.~K., {Peng}, Z.~K., {et~al.} 2021, Nature Astronomy,
  5, 911

\end{thebibliography}

\begin{appendix} 
\section{Additional material}

\begin{figure}
\centering
\includegraphics[scale=0.36]{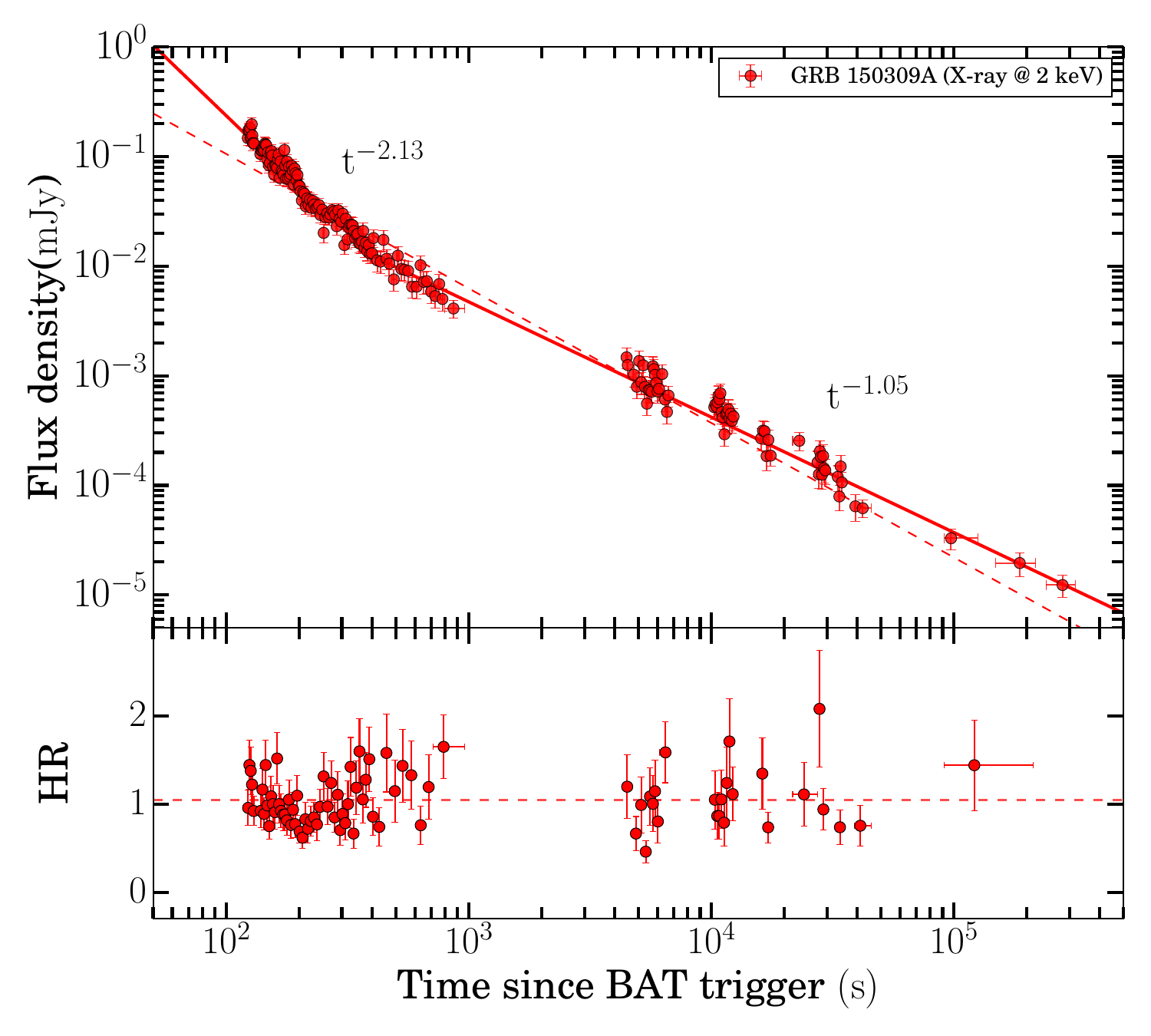}
\caption{{\bf X-ray flux light curve of \thisgrb.} Top panel shows the XRT flux light curves at 2 \keV energy range. The light curve has been fitted with power-law and broken power-law (best fit) models. Bottom panel: Evolution of hardness ratio in the XRT energy channel. The vertical red dashed line indicates the mean value of hardness ratio for \thisgrb.}
\label{fig:xrtlc}
\end{figure}

\begin{table*}
\centering
\begin{scriptsize}
\caption{Time-resolved spectral fitting of the first episode of \thisgrb using \sw{Band} and \sw{CPL} models. The listed flux values are estimated in 8 \keV-40 MeV.} 
\begin{tabular}{|c|c|c|ccccc|cccc|c|}
\hline
T$_{\rm start}$ (s) & T$_{\rm stop}$ (s) & Significance & \boldmath $\alpha_{\rm pt}$ & \boldmath $\beta_{\rm pt}$ & \boldmath \Ep (\keV) &  \bf (Flux $\times 10^{-06}$)  & \bf DIC$_{\rm Band}$ & \boldmath $\Gamma_{\rm CPL}$ & \boldmath $E_{\rm c}$ (\keV) &  \bf (Flux $\times 10^{-06}$)  & \bf DIC$_{\rm CPL}$ & \rm \bf $\Delta$ DIC\\ \hline
2.119 & 9.975 & 40.59 & $-0.15_{-0.08}^{+0.08}$ & $-3.07_{-0.23}^{+0.22}$ & $114.09_{-4.54}^{+4.34}$ & 0.58 & 4572.74 & $-0.23_{-0.07}^{+0.07}$ & $68.02_{-4.67}^{+4.52}$ & 0.49 & 4570.91 & 1.83 \\
9.975 & 11.875 & 33.09 & $-0.36_{-0.10}^{+0.10}$ & $-2.63_{-0.25}^{+0.24}$ & $149.37_{-11.59}^{+11.85}$ & 1.37 & 2593.98 & $-0.48_{-0.08}^{+0.08}$ & $112.10_{-12.12}^{+11.79}$ & 1.02 & 2593.37 & 0.61 \\
11.875 & 16.349 & 74.58 & $-0.38_{-0.05}^{+0.05}$ & $-2.44_{-0.11}^{+0.12}$ & $172.78_{-8.02}^{+7.96}$ & 2.8 & 3898.49 & $-0.50_{-0.03}^{+0.03}$ & $134.78_{-6.69}^{+6.73}$ & 1.87 & 3922.39 & -23.9 \\
16.349 & 20.547 & 82.42 & $-0.54_{-0.04}^{+0.04}$ & $-2.63_{-0.17}^{+0.17}$ & $215.39_{-8.90}^{+8.98}$ & 3.12 & 3731.73 & $-0.60_{-0.03}^{+0.03}$ & $168.88_{-8.17}^{+8.03}$ & 2.37 & 3741.33 & -9.6 \\
20.547 & 21.861 & 33.15 & $-0.51_{-0.09}^{+0.08}$ & $-2.63_{-0.25}^{+0.25}$ & $176.76_{-14.07}^{+14.22}$ & 1.95 & 2035.34 & $-0.60_{-0.07}^{+0.07}$ & $142.79_{-16.29}^{+16.30}$ & 1.47 & 2036.18 & -0.84 \\
21.861 & 23.799 & 48.52 & $-0.61_{-0.06}^{+0.06}$ & $-2.31_{-0.16}^{+0.16}$ & $198.38_{-16.56}^{+16.64}$ & 3.15 & 2656.74 & $-0.71_{-0.05}^{+0.05}$ & $185.58_{-17.42}^{+17.54}$ & 1.93 & 2671.61 & -14.87 \\
23.799 & 25.994 & 34.11 & $-0.77_{-0.07}^{+0.08}$ & $-2.75_{-0.33}^{+0.34}$ & $187.26_{-18.14}^{+18.09}$ & 1.3 & 2701.82 & $-0.81_{-0.06}^{+0.06}$ & $169.40_{-21.11}^{+21.18}$ & 1.03 & 2700.68 & 1.14 \\
25.994 & 29.582 & 29.29 & $-0.75_{-0.09}^{+0.09}$ & $-2.50_{-0.28}^{+0.28}$ & $161.17_{-17.65}^{+18.08}$ & 0.87 & 3385.32 & $-0.82_{-0.08}^{+0.08}$ & $157.01_{-24.01}^{+23.98}$ & 0.6 & 3387.92 & -2.6 \\
31.088 & 60.001 & 18.84 & $-0.95_{-0.15}^{+0.15}$ & $-2.83_{-0.33}^{+0.33}$ & $87.04_{-11.39}^{+11.25}$ & 0.12 & 6215.44 & $-1.03_{-0.12}^{+0.12}$ & $100.06_{-21.18}^{+21.49}$ & 0.09 & 6212.99 & 2.45 \\ \hline
\end{tabular}
\end{scriptsize}
\label{TRS_Table_Bayesian}
\end{table*}

\begin{figure*}
\centering
\includegraphics[scale=0.25]{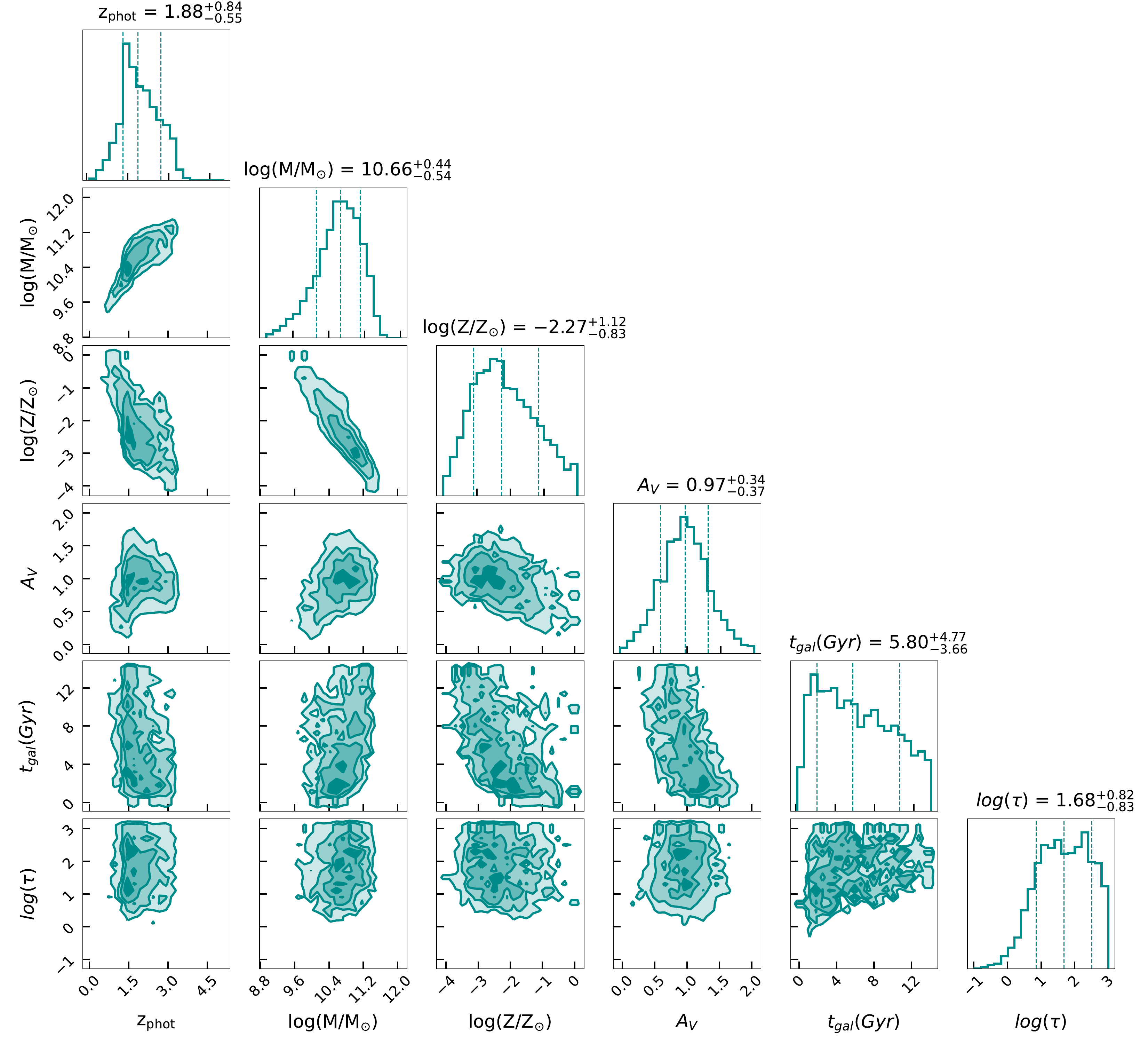}
\caption{The posterior distributions for the SED model parameters of the potential host galaxy of \thisgrb obtained using nested sampling via \sw{dynesty} using \sw{Prospector} software.}
\label{PotentialHG-SED_corner}
\end{figure*}	   

\end{appendix}

\end{document}